\documentclass[10pt,twocolumn,showpacs,showkeys,preprintnumbers,amssymb,aps,superscriptaddress,pre]{revtex4-2}
\usepackage{color}
\usepackage{amsmath, amssymb}
\usepackage{graphicx}
\usepackage{bm}
\usepackage{subfigure}
\usepackage{ar}

\begin{document}

\preprint{APS/123-QED}
\title{Magnetization plateaus and enhanced magnetocaloric effect of a spin-1/2 Ising-Heisenberg and Heisenberg double sawtooth ladder with four-spin interaction}

\author{Hamid Arian Zad}
\affiliation{A.I. Alikhanyan National Science Laboratory, 0036, Yerevan, Armenia}
\affiliation{Department of Theoretical Physics and Astrophysics, Faculty of Science, P. J. S\v{a}f{\' a}rik University, Park Angelinum 9, 041 54 Ko\v{s}ice, Slovak Republic}
\author{Vadim Ohanyan}
\affiliation{Laboratory of Theoretical Physics, Yerevan State University, Alex Manoogian 1, 0025 Yerevan, Armenia}
\affiliation{CANDLE, Synchrotron Research Institute, 31 Acharyan Str., 0040 Yerevan, Armenia}
\author{Azam Zoshki}
\affiliation{Department of Theoretical Physics and Astrophysics, Faculty of Science, P. J. S\v{a}f{\' a}rik University, Park Angelinum 9, 041 54 Ko\v{s}ice, Slovak Republic}
\author{Jozef Stre\v{c}ka}
\affiliation{Department of Theoretical Physics and Astrophysics, Faculty of Science, P. J. S\v{a}f{\' a}rik University, Park Angelinum 9, 041 54 Ko\v{s}ice, Slovak Republic}

\date{\today}

\begin{abstract}
The ground state, the entropy and the magnetic Gr{\" u}neisen parameter of the antiferromagnetic spin-1/2 Ising-Heisenberg model on a double sawtooth ladder are rigorously investigated using the classical transfer-matrix technique.
The model includes the XXZ interaction between the interstitial Heisenberg dimers, the Ising coupling between nearest-neighbor spins of the  legs and rungs, and additional cyclic four-spin Ising term in each square plaquette.
For a particular value of the cyclic four-spin exchange we found in the ground-state phase diagram of the Ising-Heisenberg ladder a quadruple point, at which four different ground states coexist together.
 During an adiabatic demagnetization process a fast cooling accompanied with an enhanced magnetocaloric effect can be detected
nearby this quadruple point.
The ground-state phase diagram of the Ising-Heisenberg ladder is confronted with the zero-temperature magnetization process of the purely quantum Heisenberg ladder, which is calculated by using exact diagonalization (ED) based on the Lanczos algorithm for a finite-size ladder of 24 spins and the density-matrix renormalization group (DMRG) simulations for a finite-size ladder with up to 96 spins. Some indications of existence of intermediate magnetization plateaus in the magnetization process of the full Heisenberg model for a small but non-zero four-spin Ising coupling were found. The DMRG results reveal that the quantum Heisenberg double sawtooth ladder exhibits a quantum Luttinger spin-liquid phase that is absent in the Ising-Heisenberg counterpart model. Except this difference the magnetic behavior of the full Heisenberg model is quite analogous to its simplified Ising-Heisenberg counterpart and hence, one may bring insight into the fully quantum Heisenberg model from rigorous results for the Ising-Heisenberg model.

\end{abstract}

\maketitle

\section{Introduction} \label{sec:level1}
The adiabatic change of the temperature of magnetic materials under the variation of an external magnetic field is known as a magnetocaloric effect (MCE). The adiabatic demagnetization is the main ingredient of the magnetic refrigeration technique, whereby the enhanced MCE has important application at a room-temperature as well as low-temperature refrigeration (for recent reviews on the subject see Refs. \cite{oja97,gsch05}). Another important application of the MCE is the use of the magnetocaloric anomalies at the magnetic phase transitions for the obtaining of phase diagrams of the magnetic materials in magnetic field-temperature plane.  The standard quantity to characterize the MCE is the so-called adiabatic cooling rate, defined as
\begin{equation}\label{SLhamiltonian}
\begin{array}{lcl}
\Gamma=\frac{1}{T}\big(\frac{\partial T}{\partial B}\big)_S=-\frac{T}{\mathcal{C}_B}\big(\frac{\partial S}{\partial B}\big)_T=-\frac{T}{\mathcal{C}_B}\big(\frac{\partial \mathcal{M}}{\partial T}\big)_B,
\end{array}
\end{equation}
where $\mathcal{C}_B$ is the heat capacity at the constant magnetic field, $T$ is the temperature and $B$ is the applied magnetic field. In the recent decades a series of  theoretical researches have been conducted on the MCE in low-dimensional quantum and Ising spin models \cite{zhit04,hon09,Honecker2009,Vadim2010,Vadim2012EPJB,gal14,str15,str17a,kar17,zuk18,Beckmanna2018, Vadim2012, str14,gal16,tor16,ale18,gal18}. The exact results obtained within the low-dimensional quantum and mixed quantum-classical interacting spin models figure out many important features of the MCE, particularly the enhancing role of frustration and residual entropy, the possibility of magnetic cooling or magnetic heating during the adiabatic demagnetization, a deep connection of the MCE and quantum phase transitions, etc. Besides the investigation of the behaviour of adiabatic cooling rate, important information about the MCE can be obtained form the plots of the isentropes in the temperature-magnetic field plane.

In recent years, a number of exact results on the MCE in one-dimensional Ising-Heisenberg  spin models have been obtained \cite{Vadim2012,str14,gal16,tor16,ale18,gal18}. The common feature of these models is a regular alternation of small clusters of quantum Heisenberg spins with classical Ising spins in such a way that local Hamiltonians for individual blocks commute with each other. This specific condition allows one to obtain an exact solution of the Ising-Heisenberg spin chains in terms of the generalized classical transfer-matrix method. Although the regular alternation of the Heisenberg and Ising spins rarely occurs in real magnetic materials, a few notable examples of the exactly solved Ising-Heisenberg spin chains have remarkably accurately elucidated the magnetic properties of specific magnetic compounds such as Cu(3-Clpy)$_2$(N$_3$)$_2$ \cite{str05}, [(CuL)$_2$Dy][Mo(CN)$_8$] \cite{heu10,bel14}, [Fe(H$_2$O)(L)][Nb(CN)$_8$][Fe(L)] \cite{sah12}, Dy(NO$_3$)(DMSO)$_2$Cu(opba)(DMSO)$_2$ \cite{str12,han13}, [CuMn(L)][Fe(bpb)(CN)$_{2}$] $\cdot$ ClO$_{4}$ $\cdot$ H$_{2}$O \cite{sou20} and [Dy(hfac)$_2$(CH$_3$OH)]$_2$[Cu(dmg)(Hdmg)]$_2$ \cite{str20,gal22}.

A few comments are in order as far the four-spin Ising coupling is concerned, which can be alternatively considered as the Ising limit of the cyclic permutation of the quantum spins or higher-order exchange process. As a matter of fact, these exchange terms are proven to be of ultimate importance for understanding of magnetic properties of the solid He$^3$ \cite{roger} as well as of some cuprates (see Ref. \cite{Muller} and references therein). The Ising and mixed Ising-Heisenberg spin models with four-spin Ising coupling (product of four Ising variables) have been also considered in several papers as a Ising limit of the four-spin cyclic permutation as well as in an independent setups \cite{gal14,ara03,oha05,ana07,hov09,ana12,gal13,Arian1,Arian2,Arian3}. The Ising-Heisenberg model on the sawtooth chain \cite{Ohanyan2009,Bellucci,bel13}, on the two-leg ladder \cite{taras13,jozef2020} and on the tetrahedral chain \cite{jozef2014} has been the topic of earlier studies.

On the one hand, both spin-1/2 and mixed spin-(1/2,1) Ising-Heisenberg double sawtooth ladders with the cyclic four-spin interaction have been exactly solved within the modified classical transfer-matrix technique and some results for their ground-state phase diagram, magnetic and thermodynamic properties have been reported in previous works \cite{Arian1,Arian2,Arian3}.
 Moreover, the ground-state phase diagram of the spin-1/2 Heisenberg model defined on a two-leg ladder with alternating rung exchanges in the absence of external field has been studied in Ref. \cite{Amiri2015}. The authors found rich ground-state phase diagram for a more general case of the full Heisenberg model on a honeycomb-like ladder including rung-singlet, saturated ferromagnetic states and a ferrimagnetic Luttinger spin-liquid phase. Motivated by the fascinating behavior observed in ladder-type models, extensive research has been conducted to understand the magnetic excitations and ground-state phase diagram of the frustrated spin ladder with next-nearest-neighbor interactions \cite{Sugimoto2013}.

\begin{figure}
  \centering
  \resizebox{0.5\textwidth}{!}{
 \includegraphics[trim=70 540 100 120, clip]{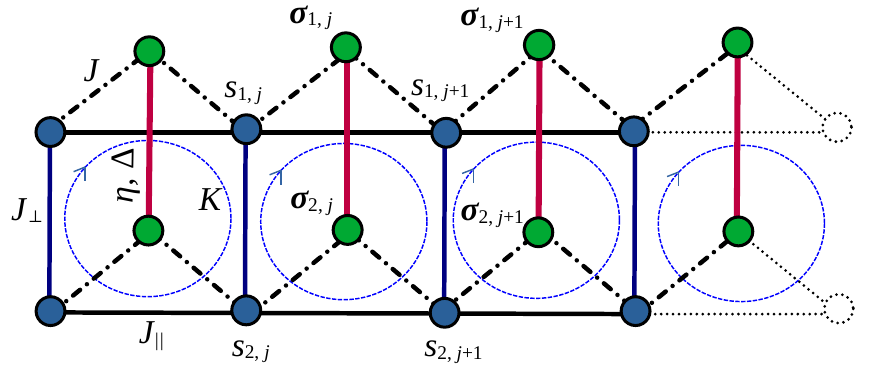} 
}
\vspace{-0.75cm}
\caption{A schematic illustration of the magnetic structure of the frustrated spin-1/2 double sawtooth ladder. The balls denote spin-1/2 particles. Green balls represent Heisenberg spins that are connected with each other with red solid lines. The dark-blue balls show Ising spins. The blue dashed circle in each square plaquette denotes four-spin Ising coupling.}
\label{fig:SpinLadder}
\end{figure}

The frustrated double sawtooth ladder shown in Fig.  \ref{fig:SpinLadder} is an attractive spin model since it could result in a controversy regarding a toy model of two-leg ladder with the next-nearest-neighbor interaction
that exhibits exotic magnetic states.
 The later property originates from the geometric structure of the model which links two frustrated sawtooth spin
chains with each other through different rungs \cite{Koteswararao2007,Mentr2009,Koteswararao2010}.

Although previous studies uncovered large number of exciting magnetic properties of a set of similar quantum spin ladder models, some important insights still remain controversial. For example, the zero-temperature magnetization, MCE and cooling/heating process of the double sawtooth ladder with a small but nonzero four-spin Ising interaction still remain unresolved issues to deal with. In the present paper, we therefore aim to investigate the MCE of the Ising-Heisenberg double sawtooth ladder particularly close to a critical cyclic four-spin Ising interaction at which four ground states of this model coexist together. Moreover, magnetization process of the full Heisenberg double sawtooth ladder is investigated using full ED method. We use the QuSpin package \cite{Weinberg2017,Weinberg2019} to apply ED for solving finite-size ladder of 24 spins. Besides, ITensor package \cite{Miles2022} is used to perform DMRG method based on tensor network algorithms for diagonalizing large-size ladder with 64, 80 and 96 number of spins with conserving total quantum number $S^{z}_\text{T}$. We particularly focus on the influences of the Heisenberg exchange anisotropy and four-spin Ising coupling on the magnetization behavior of the full Heisenberg double sawtooth ladder.

The organization of this paper is as follows: in the next section we describe in detail the spin-1/2 Ising-Heisenberg model on the double sawtooth ladder and the analytical procedure to solve this model.
Then, we provide a full description of the ground-state phase diagram and critical behavior of the Ising-Heisenberg model. Furthermore, the behavior of entropy and the Gr{\" u}neisen parameter with respect to the magnetic field and temperature for various fixed values of the model parameters are discussed. Effects of the four-spin Ising coupling on the adiabatic demagnetization process of the Ising-Heisenberg model are analyzed, as well. In Sec. \ref{fullHeisenberg}, we introduce the full quantum Heisenberg model on the frustrated spin-1/2 double sawtooth ladder. Next, we demonstrate our numerical methods, i.e., the ED method based on the Lanczos algorithm and the DMRG to calculate zero-temperature magnetization process of the full Heisenberg double sawtooth ladder. Sec. \ref{conclusion} contains a brief conclusion.

\section{Ising-Heisenberg double sawtooth ladder}\label{Ising-Heisenberg Model}
Let us consider a frustrated spin-1/2  Ising-Heisenberg double sawtooth ladder (or decorated two-leg ladder) with four-spin Ising coupling as schematically illustrated in Fig. \ref{fig:SpinLadder}.
The total Hamiltonian of this model is given by the following expression:
\begin{eqnarray}\label{SLhamiltonian}
H=
&& \sum\limits_{j=1}^N\Big[\left({\boldsymbol \sigma}_{1,j}\cdot{\boldsymbol \sigma}_{2,j}\right)_{\eta,\Delta}+
\nonumber \\
&& \sum\limits_{a=1,2}\big({J}_{\parallel}{s}_{a,j}{s}_{a,j+1}+J \sigma^z_{a,j}({s}_{a,j}+{s}_{a,j+1})\big)
\nonumber \\
&& +{J}_{\perp}{s}_{1,j}{s}_{2,j}+Ks_{1,j}s_{1,j+1}s_{2,j}s_{2,j+1}
\nonumber\\
&& -B\sum\limits_{a=1,2}\left(\sigma_{a,j}^z+s_{a,j}\right)\Big],
\end{eqnarray}
where $N$ denotes the total number of unit cells, $s_{a,j}$ are Ising spin variables taking values $1$ or $-1$, $B$ is the external magnetic field applied along the $z$-direction and each pair of quantum spins ${\boldsymbol \sigma}_{1,j}$ and ${\boldsymbol \sigma}_{2,j}$ interacts via the XXZ exchange coupling defined through the Pauli matrices ${\sigma}_i^\alpha$($\alpha=x,y,z$) as follows:
\begin{equation}\label{Four Spin}
\begin{array}{lcl}
 \left({\boldsymbol \sigma}_{1,j}\cdot{\boldsymbol \sigma}_{2,j}\right)_{\eta,\Delta}=\big[\eta\big({\sigma}^{x}_{1,j}{\sigma}^{x}_{2,j}
 +{\sigma}^{y}_{1,j}{\sigma}^{y}_{2,j}\big)+\Delta{\sigma}^{z}_{1,j}{\sigma}^{z}_{2,j}\big].
\end{array}
\end{equation}
The coupling constants ${J}_{\perp}$ and ${J}_{\parallel}$ are the Ising-type couplings on the rungs and along the legs, respectively, and the coupling constant $K$ accounts for the four-spin Ising term in each square plaquette composed from four Ising spins belonging to two neighboring unit blocks. The coupling constant $J$ is the Ising coupling between the Ising spins on the legs and two quantum spins of the interstitial Heisenberg dimer on the tips. This coupling is supposed to include only $z$-component of the quantum spins. Let us mention that all parameters here are supposed to be dimensionless taking $J_{\parallel}=J_{\perp}$ as energy unit.

\subsection{The exact solution in terms of the generalized classical transfer-matrix method}
Let us present a few crucial steps of the analytical procedure used to calculate the partition function $Z$ of the model within the generalized classical transfer-matrix technique:
\begin{eqnarray}
&& Z=\sum_{\left(s_1\right)}\sum_{\left(s_2\right)}\mbox{Tr}_{\left({\boldsymbol \sigma}_{1}, {\boldsymbol \sigma}_{2}\right)} e^{-\beta H}=\nonumber \\
&& \sum_{\left(s_1\right)}\sum_{\left(s_2\right)} \prod_{j=1}^N\mathbf{T}\left(s_{1,j},s_{2,j}|s_{1,j+1}, s_{2,j+1}\right)=\mbox{Tr}\;\mathbf{T}^N,
\end{eqnarray}
where $4\times 4$ transfer matrix is given by
\begin{eqnarray}
&&\mathbf{T}\left(s_{1,j},s_{2,j}|s_{1,j+1}, s_{2,j+1}\right) \nonumber \\
&&=\exp\left\{-\beta H^\text{I}\left(s_{1,j},s_{2,j}|s_{1,j+1}, s_{2,j+1}\right)\right\} \times \nonumber\\
&& \quad W\left(s_{1,j},s_{2,j}|s_{1,j+1}, s_{2,j+1}\right),
\end{eqnarray}
where the Ising part of the system's Hamiltonian reads
\begin{eqnarray}
&&H^\text{I}\left(s_{1,j},s_{2,j}|s_{1,j+1}, s_{2,j+1}\right)={J}_{\parallel}\sum\limits_{a=1,2}{s}_{a,j}{s}_{a,j+1}\nonumber \\
&&+{J}_{\perp}{s}_{1,j}{s}_{2,j}+K s_{1,j}s_{1,j+1}s_{2,j}s_{2,j+1}-B\sum\limits_{a=1,2}s_{a,j},\nonumber\\
\end{eqnarray}
and the Boltzmann factor of the quantum dimer, $W$, is
\begin{eqnarray}
W\left(s_{1,j},s_{2,j}|s_{1,j+1}, s_{2,j+1}\right)=\sum_{l=1}^{4}e^{-\beta\varepsilon_n\left(s_{1,j},s_{2,j}|s_{1,j+1}, s_{2,j+1}\right)},
\nonumber
\end{eqnarray}
in which $\varepsilon_n\left(s_{1,j},s_{2,j}|s_{1,j+1}, s_{2,j+1}\right)$ with $n=1,...,4$ are the eigenvalues of the quantum part of the Hamiltonian:
\begin{eqnarray}\label{Hq}
H^\text{q}=\left({\boldsymbol \sigma}_{1,j}\cdot{\boldsymbol \sigma}_{2,j}\right)_{\eta,\Delta}-\sum\limits_{a=1,2}\left[B-J( {s}_{a,j}+{s}_{a,j+1})\right]\sigma^z_{a,j}.\nonumber\\
\end{eqnarray}
The eigenvalues and eigenvectors of the quantum part of the total Hamiltonian (\ref{SLhamiltonian}) explicitly depend on the values of four Ising spin variables, interacting with the quantum spin dimer. They can be easily found by the straightforward diagonalization of the quantum Hamiltonian (\ref{Hq}) in the standard Ising basis
$\lbrace \vert\!\uparrow\uparrow\rangle,\vert\!\uparrow\downarrow\rangle,\vert\!\downarrow \uparrow\rangle,\vert\!\downarrow \downarrow\rangle \rbrace$. The eigenvalues are
\begin{eqnarray}
&& \varepsilon_{1,2}=-{\Delta}\pm  \sqrt{J^2\left(s_{1,j}-s_{2,j}+s_{1,j+1}-s_{2,j+1}\right)^2+\eta^2},\nonumber \\
&& \varepsilon_{3,4}={\Delta}\pm \left[2B-J(s_{1,j}+s_{2,j}+s_{1,j+1}+s_{2,j+1})\right],
\end{eqnarray}
and the corresponding eigenvectors are given by
\begin{eqnarray}
&& \vert\varphi_{1,2}\rangle=\frac{1}{\sqrt{1+A_{\pm}^2}}\left(\vert\!\uparrow\downarrow\rangle+A_{\pm}\vert\!\downarrow\uparrow\right),\\
&& \vert\varphi_3\rangle=\vert\!\downarrow\downarrow\rangle, \nonumber \\
&& \vert\varphi_4\rangle=\vert\!\uparrow\uparrow\rangle, \nonumber
\end{eqnarray}
where
\begin{eqnarray}
&& A_{\pm}=\dfrac{1}{\eta}\Big[ J\left(s_{2,j}-s_{1,j}+s_{2,j+1}-s_{1,j+1}\right)\pm \nonumber\\
&& \sqrt{J^2\left(s_{2,j}-s_{1,j}+s_{2,j+1}-s_{1,j+1}\right)^2+\eta^2}\Big].
\end{eqnarray}
Thus, we see that the eigenvectors for the quantum spin dimers  depend on the four values of the Ising spins they are interacting with.
The explicit form of the $4\times 4$ classical transfer matrix is:
\begin{widetext}
\begin{eqnarray}
\mathbf{T} = \left(\begin{array}{lccr}
 z_1^{-1}z_2^{-1}\lambda^{-1}\mu^2(\chi_0+\psi_2) & z_2^{-1}\lambda\mu^2(\chi_1+\psi_1) & z_2^{-1}\lambda\mu^2(\chi_1+\psi_1)  & z_1z_2^{-1}\lambda^{-1}\mu^2(\chi_0+\psi_0) \\
 z_2\lambda(\chi_1+\psi_1) & z_1z_2^{-1}\lambda(\chi_2+\psi_0) & z_1z_2\lambda^{-1}(\chi_0+\psi_0) & z_2\lambda(\chi_1+\psi_{-1}) \\
 z_2\lambda(\chi_1+\psi_1) & z_1z_2\lambda^{-1}(\chi_0+\psi_0) & z_1^{-1}z_2\lambda^{-1}(\chi_2+\psi_0) & z_2\lambda(\chi_1+\psi_{-1}) \\
 z_1z_2^{-1}\lambda^{-1}\mu^{-2}(\chi_0+\psi_0) & z_2^{-1}\lambda^{-1}\mu^{-2}(\chi_1+\psi_{-1}) & z_2^{-1}\lambda^{-1}\mu^{-2}(\chi_1+\psi_{-1}) &  z_1^{-1}z_2^{-1}\lambda^{-1}\mu^{-2}(\chi_0+\psi_{-2})
 \end{array}\right),
\end{eqnarray}
\end{widetext}
where the following notations are adopted:
\begin{eqnarray}
&& z_1=e^{2\beta J_{\parallel}},\; z_2=e^{\beta J_{\perp}},\; \lambda=e^{\beta K},\; \mu=e^{\beta B}, \\
&& \chi_n=2e^{\beta\Delta}\cosh\frac{\beta}{2}\sqrt{(2nJ)^2+\eta^2}, \;\; n=0, 1, 2. \nonumber \\
&& \psi_n=2e^{-\beta\Delta}\cosh\left(\beta\left(B-n J\right)\right), \;\; n=-2, -1, 0, 1, 2. \nonumber
\end{eqnarray}

\subsection{Ground states}
\begin{figure*}
  \centering
 \resizebox{1\textwidth}{!}{
 \includegraphics[trim=20 550 25 90, clip]{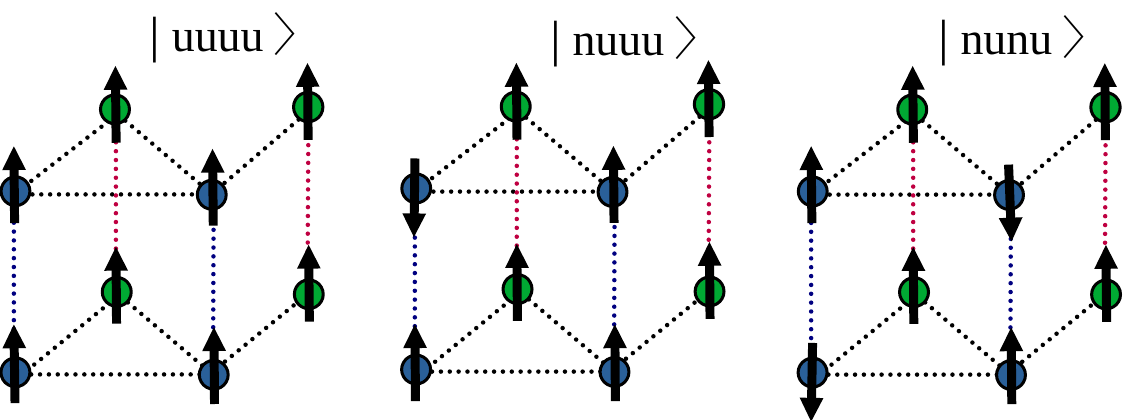}
  \includegraphics[trim=20 350 25 300, clip]{SpinConfig}
 \includegraphics[trim=200 150 200 500, clip]{SpinConfig}

}
\vspace{-0.25cm}
\caption{The spin configurations of the spin-1/2 double sawtooth ladder associated to the possible ground states (\ref{Eigns}).
Each panel indicates spin arrangement of two successive unit blocks that are repeated throughout the ladder. Tick arrows represent the spin orientations, whereas oval shapes stand for the Heisenberg dimers in quasi-singlet state $|\varphi_2 \rangle$. Notations in the kets denote the orientation of the two spins on each rung such that "up" means both spins of the same rung are up, while "n" indicates that one spin of the rung is up and another one is down, and "s" means that the state of the two spins on the identical rung is quasi-singlet state $|\varphi_2 \rangle$.}
\label{fig:L_SpinArrangs1}
\end{figure*}

The system under consideration possesses several ground states, depending on its microscopic parameters,
$\eta$; $\Delta$; $J_{\perp}$; $J$; $J_{\parallel}$; $K$ and $B$. As the main goal of this section is to
clarify the role of the Heisenberg exchange anisotropy $\Delta$ and four-spin Ising interaction $K$ in the enhancement of the
MCE of the Ising-Heisenberg double sawtooth ladder, we here focus on a peculiar range of parameters, which admit special points in the ground-state phase diagram with a high degeneracy.
The MCE is stronger around triple points and particularly around points of confluence of more phases \cite{Vadim2012EPJB}. We are going to examine the ground-state phase diagram, MCE and cooling rate of the spin-1/2 Ising-Heisenberg double sawtooth ladder with nonzero four-spin Ising interaction $K$. To provide a clear definition of the relevant ground states of the model, we demonstrate their special spin configurations in Fig. \ref{fig:L_SpinArrangs1}.

Among these ground states, $|\text{usus}\rangle$, $|\text{nsus}\rangle$, $|\text{nsns}\rangle_1$  and $|\text{nsns}\rangle_2$ include the dimeric quasi-singlet state $|\varphi_2 \rangle$ and are subject to the doubling of the unit cell, which is the
manifestation of the quantum antiferromagnetic coupling along the alternating rungs. From this point of view, we compose the spin orientation of two successive unit cells in our notation. The relevant ground
states with the corresponding magnetization values per unit cell are given by the following expressions:

\begin{equation}\label{Eigns}
\begin{array}{lcl}
\vert \mathrm{uuuu} \rangle=\displaystyle\prod_{i=1}^{N/2}\vert\! \uparrow\uparrow\rangle_{2i-1} \otimes\vert\varphi_4\rangle_{2i-1}\otimes\vert\! \uparrow\uparrow\rangle_{2i}\otimes\vert\varphi_4\rangle_{2i},\\
E_\text{uuuu}=-4B + \Delta + 2J_{\parallel} + J_{\perp} + 4J + K,\quad  M/M_s=1,\\
\vert \mathrm{nuuu} \rangle=\displaystyle\prod_{i=1}^{N/2}\vert\! \downarrow\uparrow\rangle_{2i-1} \otimes\vert\varphi_4\rangle_{2i-1}\otimes\vert\! \uparrow\uparrow\rangle_{2i}\otimes\vert\varphi_4\rangle_{2i},\\
E_\text{nuuu}=-3B + \Delta + 2J - K, \quad  M/M_s= \dfrac{3}{4},\\
\vert \mathrm{nunu} \rangle=\displaystyle\prod_{i=1}^{N/2}\vert\! \uparrow\downarrow\rangle_{2i-1} \otimes\vert\varphi_4\rangle_{2i-1}\otimes\vert\! \downarrow\uparrow\rangle_{2i}\otimes\vert\varphi_4\rangle_{2i},\\
E_\text{nunu}=-2B+\Delta-2J_{\parallel}-J_{\perp}+K, \quad  M/M_s=\dfrac{1}{2},\\
\vert \mathrm{usus} \rangle=\displaystyle\prod_{i=1}^{N/2}\vert\! \uparrow\uparrow\rangle_{2i-1} \otimes\vert\varphi_2\rangle_{2i-1}\otimes\vert\! \uparrow\uparrow\rangle_{2i}\otimes\vert\varphi_2\rangle_{2i},\\
E_\text{usus}=-2B-2\eta-\Delta+2J_{\parallel}+J_{\perp}+K, \quad  M/M_s=\dfrac{1}{2},\\
\vert \mathrm{nsus} \rangle=\displaystyle\prod_{i=1}^{N/2}\vert\! \downarrow\uparrow\rangle_{2i-1} \otimes\vert\varphi_2\rangle_{2i-1}\otimes\vert\! \uparrow\uparrow\rangle_{2i}\otimes\vert\varphi_2\rangle_{2i},\\
E_\text{nsus}=-B-\Delta-K-2\sqrt{\eta^2+J^2},  \quad  M/M_s=\dfrac{1}{4},\\
\vert \mathrm{nsns} \rangle_{1}=\displaystyle\prod_{i=1}^{N/2}\vert\! \uparrow\downarrow\rangle_{2i-1} \otimes\vert\varphi_2\rangle_{2i-1}\otimes\vert\! \downarrow\uparrow\rangle_{2i}\otimes\vert\varphi_2\rangle_{2i},\\
E_\text{nsns}^{(1)}=-2\eta-\Delta-2J_{\parallel}-J_{\perp}+K,  \quad  M/M_s=0,\\
\vert \mathrm{nsns} \rangle_{2}=\displaystyle\prod_{i=1}^{N/2}\vert\! \downarrow\uparrow\rangle_{2i-1} \otimes\vert\varphi_2\rangle_{2i-1}\otimes\vert\! \uparrow\downarrow\rangle_{2i}\otimes\vert\varphi_2\rangle_{2i},\\
E_\text{nsns}^{(2)}=-\Delta+2J_{\parallel}-J_{\perp}-2\sqrt{\eta^2+4J^2}+K,\;  M/M_s=0.\\
\end{array}
\end{equation}
Here the following principles in the notation of ground states are adopted. The unit cell in the configurations with period doubling contains four pairs of spins, each pair is coupled together by a rung. Thus, for the two spins on a rung pointing up we use \textquotedblleft u\textquotedblright, for the pair of spins pointing in opposite directions we use \textquotedblleft n\textquotedblright, and finally for the quantum quasi-singlet state $\vert \varphi_{2}\rangle$ we use \textquotedblleft s\textquotedblright.

\begin{figure}
  \centering
  \resizebox{0.4\textwidth}{!}{
 \includegraphics[trim=40 20 50 10, clip]{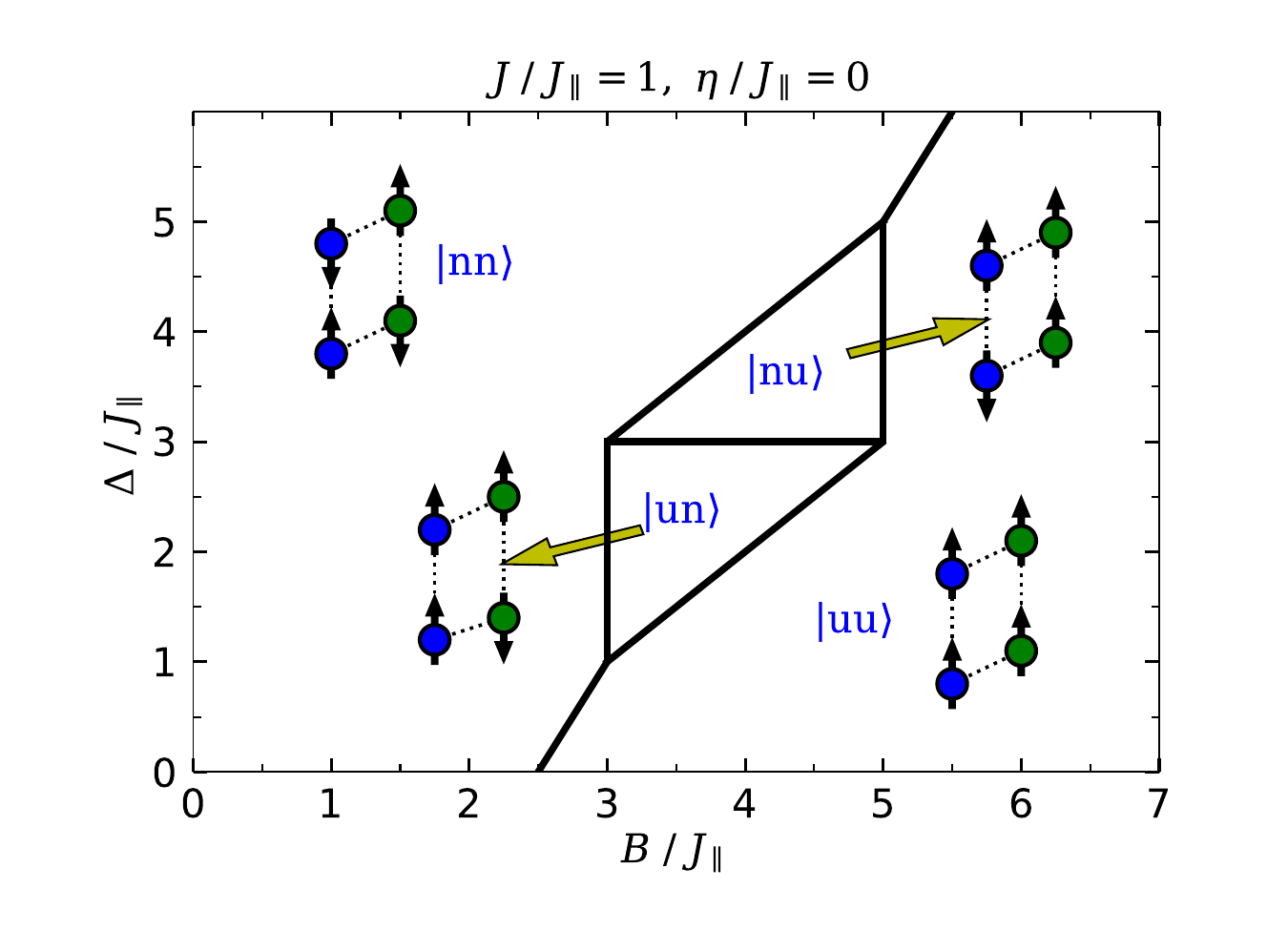}}

\caption{Ground-state phase diagram of the pure Ising double sawtooth ladder in the $\Delta/J_{\parallel}-B/J_{\parallel}$ plane for $\eta/J_{\parallel}=0$ and $J/J_{\parallel}=1$.}
\label{fig:GSPD_BDelta_Ising}
\end{figure}

For the case $\eta/J_{\parallel}=0$, where $J_{\parallel}=J_{\perp}$ is assumed as energy unit, the system is downgraded to the simple Ising model with four different ground states $\{\vert \text{nn}\rangle,\vert \text{un}\rangle,\vert \text{nu}\rangle,\vert \text{uu}\rangle\}$ with boundaries shown by solid lines in Fig. \ref{fig:GSPD_BDelta_Ising}. It can be realized from Eq. (\ref{Eigns}) that one obtains $A_{\pm}=0$ for $\eta/J_{\parallel}=0$ and $\vert \varphi_2\rangle\rightarrow\vert\!\uparrow\downarrow\rangle$ what implies that two ground states $\{\vert\text{nsns}\rangle_1,\vert \text{usus}\rangle\}$ collapse into simplified counterpart states $\{\vert\text{nn}\rangle,\vert\text{un}\rangle\}$. For better clarity we also draw in Fig. \ref{fig:GSPD_BDelta_Ising} the spin orientation of a unit block of the Ising model corresponding to each ground state. It is quite clear that here, the ground-state phase diagram of the pure Ising double sawtooth ladder is independent of the four-spin exchange coupling $K/J_{\parallel}$.

\begin{figure}
  \centering
  \resizebox{0.5\textwidth}{!}{
 \includegraphics[trim=20 25 50 50, clip]{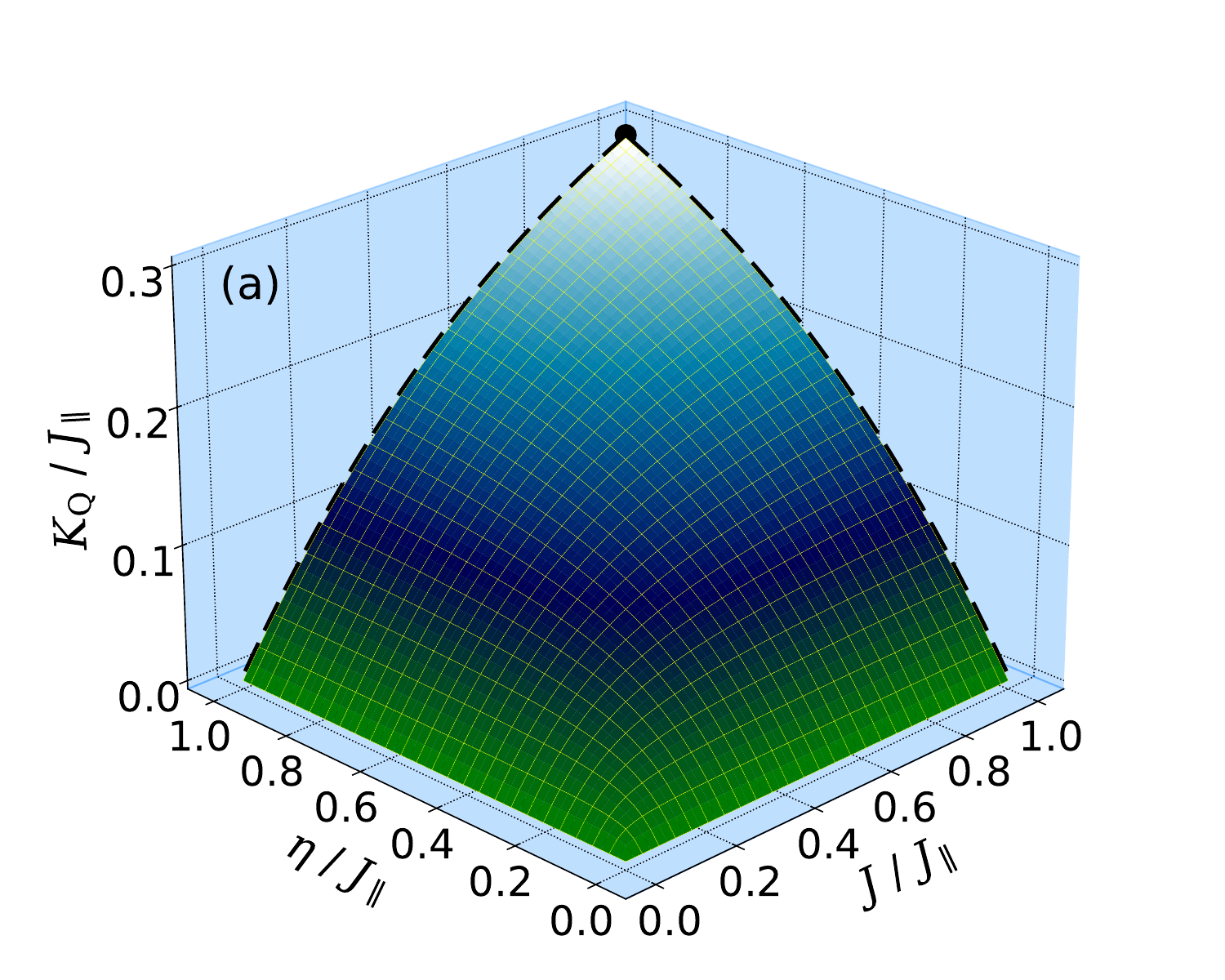}
 \includegraphics[trim=30 40 40 50, clip]{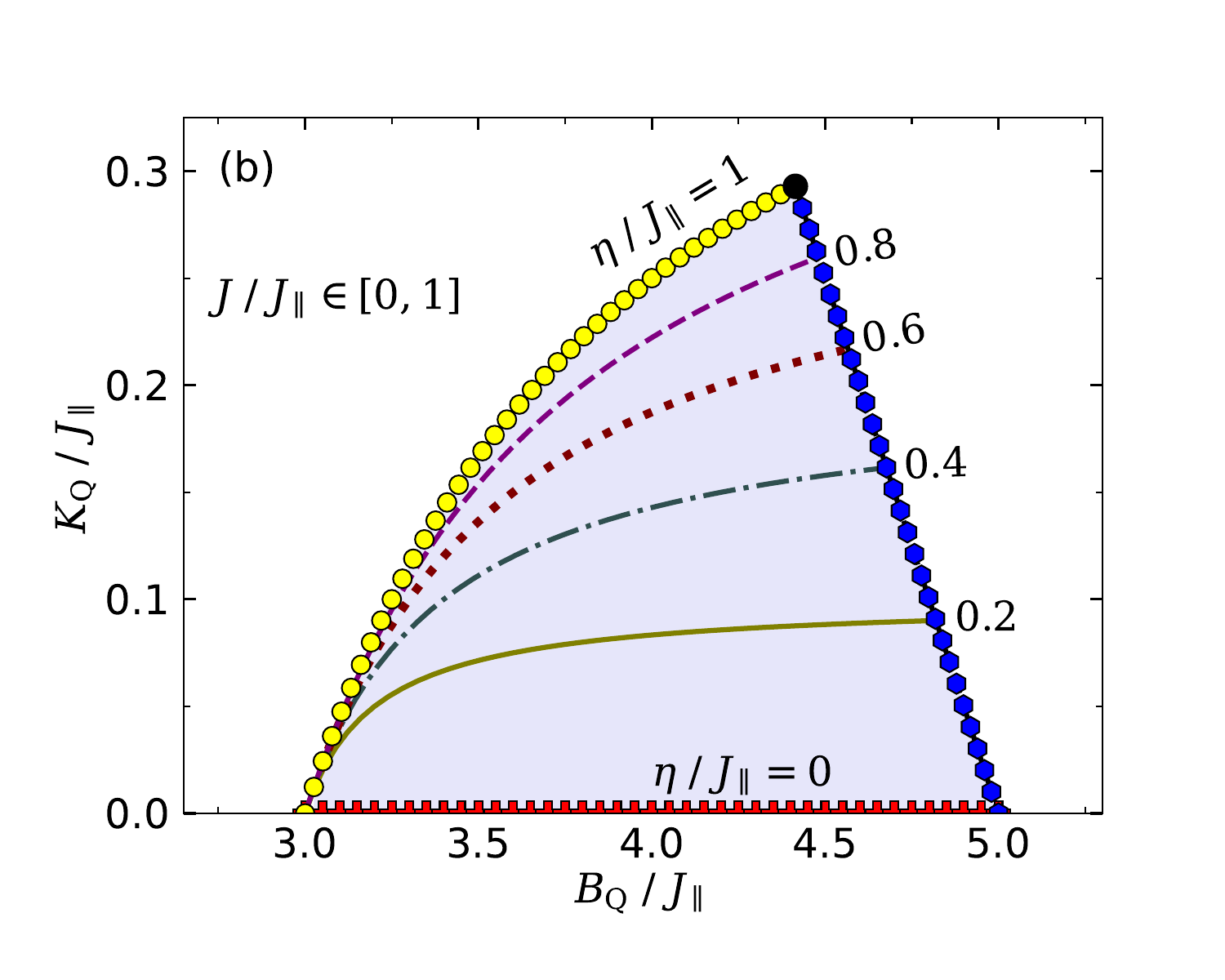}

}

\caption{(a) Topological representation of the critical four-spin exchange interaction $K_\text{Q}/J_{\parallel}=\frac{1}{2}[\eta/J_{\parallel}+J/J_{\parallel}-\sqrt{(\eta/J_{\parallel})^2+(J/J_{\parallel})^2}]$ at which four ground states $\vert\text{nsus}\rangle$, $\vert\text{usus}\rangle$, $\vert\text{nunu}\rangle$ and $\vert\text{nuuu}\rangle$ coexist together at a quadruple point with magnetic field position $B_\text{Q}/J_{\parallel}=3+2J/J_{\parallel}-2K_\text{Q}/J_{\parallel}$.
(b) The $K_\text{Q}/J_{\parallel}$ as a function of the magnetic field position of the quadruple point $B_\text{Q}/J_{\parallel}$ for $\eta/J_{\parallel}, J/J_{\parallel}\in [0,1]$. The line marked with blue hexagons shows maximum amount of $B_\text{Q}^\text{max}$ and $K_\text{Q}^\text{max}$ for $J/J_{\parallel}=1$ where $\eta/J_{\parallel}$ varies from 0 (red crosses) up to 1 (yellow circles).
$K_\text{Q}$ as a function of $B_\text{Q}$ is shown for a few selected values of $\eta/J_{\parallel}$ by lines with different styles.
Filled circle on the top of curve plotted in panel (a) shows the coordination of the point $K_\text{Q}^\text{max}/J_{\parallel}=\frac{1}{2}(2-\sqrt{2}|\eta/J_{\parallel}|)\approx 0.2929$, assuming $\eta/J_{\parallel}=J/J_{\parallel}=+1$. This point is geometrically the  same one on the ending point of the yellow curve plotted in panel (b) with $B_\text{Q}/J_{\parallel}=4.4142$.}
\label{fig:KQ}
\end{figure}

For the Ising-Heisenberg double sawtooth ladder ($\eta/J_{\parallel}\neq 0$), by simultaneously eigenenergy comparison of the four different ground states $\vert\text{nsus}\rangle$, $\vert\text{usus}\rangle$, $\vert\text{nunu}\rangle$ and $\vert\text{nuuu}\rangle$, one finds a straightforward expression for the exchange anisotropy $\Delta/J_{\parallel}$, four-spin Ising interaction $K/J_{\parallel}$ and $B/J_{\parallel}$ such that these ground states coexist together at a quadruple point.
 The two ground states $\vert\text{nunu}\rangle$ and $\vert\text{usus}\rangle$ are degenerate whenever the following condition is satisfied:
\begin{eqnarray}\label{eq:Delta_Q}
 \Delta_\text{Q}/J_{\parallel}=2 +J_{\perp}/J_{\parallel}-\eta/J_{\parallel}.
\end{eqnarray}
In our calculations we set $J_{\perp}/J_{\parallel}=\eta/J_{\parallel}=1.0$, leading to $\Delta_\text{Q}/J_{\parallel}=2.0$.
 We elucidate in Fig. \ref{fig:KQ}(a), the two-body exchange coupling dependence of the special value of the four-spin Ising interaction at the quadruple point coordinates, i.e.,
\begin{eqnarray}\label{eq:K_Q}
 K_\text{Q}/J_{\parallel}=\frac{1}{2}\Big[\eta/J_{\parallel}+J/J_{\parallel}-\sqrt{(\eta/J_{\parallel})^2+(J/J_{\parallel})^2}\Big],
\end{eqnarray}
which depends only on the $\eta/J_{\parallel}$ and $J/J_{\parallel}$. Obviously, the maximum value of $K_\text{Q}^\text{max}/J_{\parallel}=\frac{1}{2}(2-\sqrt{2}|\eta/J_{\parallel}|)$ occurs in the antiferromagnetic regime when $\eta/J_{\parallel}=J/J_{\parallel}$ (solid point on the top of curve). It is noteworthy that here, we restrict ourselves to the parameter region $0\leq\{\eta/J_{\parallel}, J/J_{\parallel}\}\leq 1$, for simplicity.
Furthermore, the magnetic field position of the quadruple point can be formulated as a function of $J/J_{\parallel}$ and $K_\text{Q}/J_{\parallel}$ as follows:
\begin{eqnarray}\label{eq:B_Q}
 B_\text{Q}/J_{\parallel}=3+2J/J_{\parallel}-2K_\text{Q}/J_{\parallel}.
\end{eqnarray}
Figure \ref{fig:KQ}(b) displays four-spin Ising interaction of the quadruple point $K_\text{Q}/J_{\parallel}$  as a function of its magnetic field position $B_\text{Q}/J_{\parallel}$ [see Eq. (\ref{eq:B_Q})] when $J/J_{\parallel}$ is varied from 0 up to 1.
It is noteworthy that the term $K_\text{Q}/J_{\parallel}$ itself depends on the $J/J_{\parallel}$ according to Eq. (\ref{eq:K_Q}). It is quite clear from this figure that by increasing the interaction ratio $J/J_{\parallel}$ the quantity $K_\text{Q}/J_{\parallel}$ increases and accordingly $B_\text{Q}/J_{\parallel}$ increases.
Lines with different styles demonstrate quantity $K_\text{Q}/J_{\parallel}$  with respect to $B_\text{Q}/J_{\parallel}$ for a few selected values of $\eta/J_{\parallel}$. All lines arise from the same minimum magnetic field point $B_\text{Q}^\text{min}/J_{\parallel}=3.0$ and terminate at different maximum magnetic field points. Regarding this, Eqs. (\ref{eq:K_Q}) and (\ref{eq:B_Q}) hold for each point of the shaded area of Fig. \ref{fig:KQ}(b) where the coordinates of the quadruple point does not exceed this area. 
The line marked with blue hexagons manifests a descending behavior of $B_\text{Q}^\text{max}/J_{\parallel}$ with respect to the interaction ratio $\eta/J_{\parallel}$. In other words, when the exchange interaction ratio $\eta/J_{\parallel}$ increases the term $B_\text{Q}^\text{max}/J_{\parallel}$ moves towards lower values, while $K_\text{Q}^\text{max}/J_{\parallel}$ increases.

\begin{figure}
  \centering
  \resizebox{0.4\textwidth}{!}{
 \includegraphics[trim=20 20 40 10, clip]{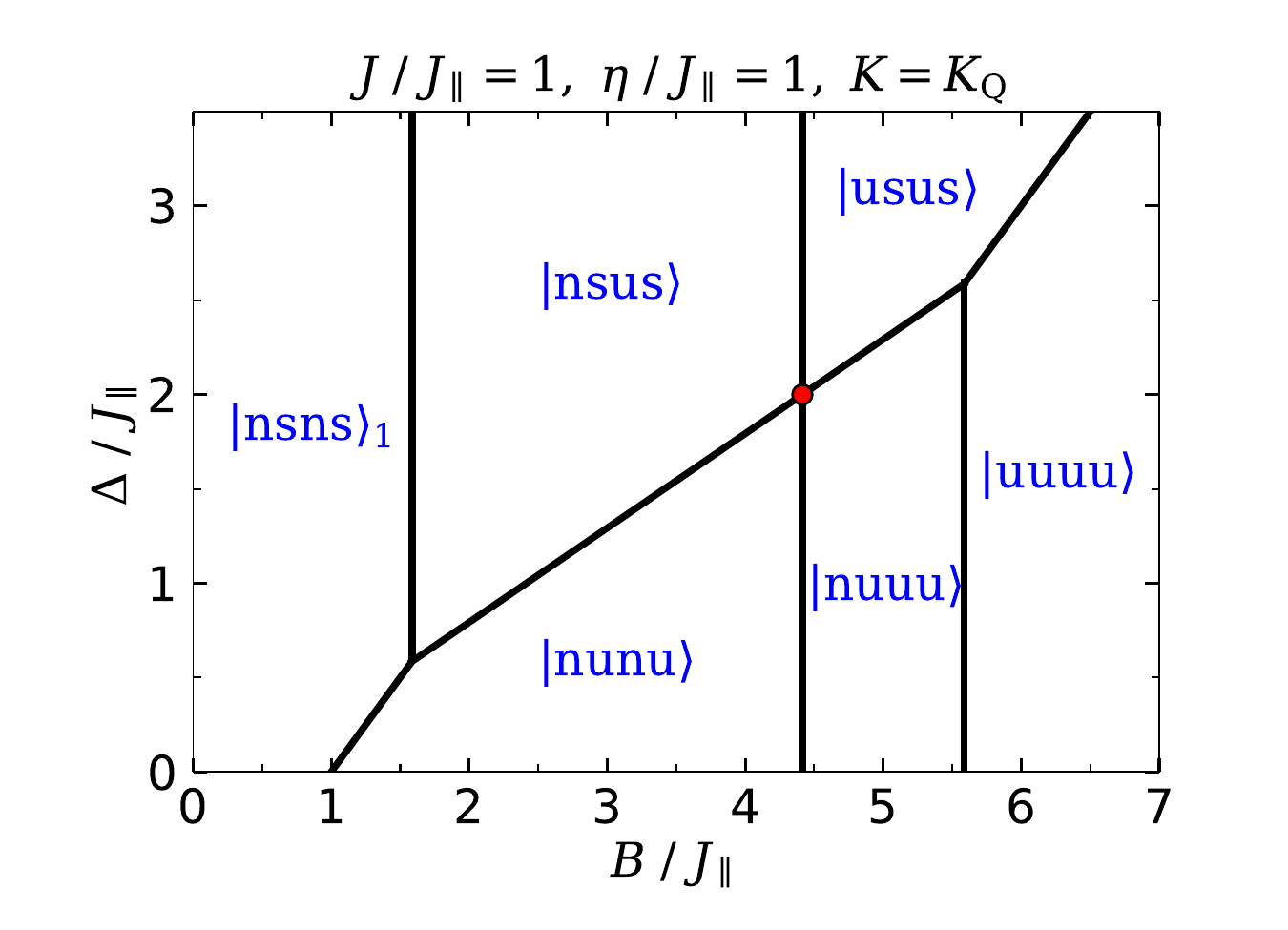}
}

\caption{Ground-state phase diagram of the Ising-Heisenberg double sawtooth ladder in the $\Delta/J_{\parallel}-B/J_{\parallel}$ plane for
$\eta/J_{\parallel}=J/J_{\parallel}=1$ and $K=K_\text{Q}\approx 0.2929$, where the quadruple point occurs at $[\Delta_\text{Q}/J_{\parallel}=2.0, B_\text{Q}/J_{\parallel}\approx 4.4142]$. Red circle manifests the quadruple point.}
\label{fig:GSPD_BDelta}
\end{figure}

Keeping this fact in mind, we plot in
 Fig. \ref{fig:GSPD_BDelta} the ground-state phase diagram of the Ising-Heisenberg ladder for $\eta/J_{\parallel}=J/J_{\parallel}=1.0$ and $K=K_\text{Q}$.
 To have a direct conjunction between Fig. \ref{fig:KQ}(b) and Fig. \ref{fig:GSPD_BDelta}, we have marked a particular point with a black circle on the upper bound of yellow curve at
  $[K_\text{Q}^\text{max}/J_{\parallel}, B_\text{Q}^\text{max}/J_{\parallel}]\approx [0.2929, 4.4142]$ that indicates the position of the quadruple point appeared in Fig. \ref{fig:GSPD_BDelta}. For higher interaction ratio $J/J_{\parallel}$, four aforementioned ground states coexist together at higher magnetic field and higher four-spin Ising interaction (see a black circle on the yellow curve in Fig. \ref{fig:KQ}(b)). Moreover, there are two triple points at $[\Delta/J_{\parallel}, B/J_{\parallel}]\approx[0.595, 1.5858]$ and at $[\Delta/J_{\parallel}, B/J_{\parallel}]\approx[2.595, 5.5858]$ where three different ground states coexist together.

As a result, the position of quadruple point moves towards higher magnetic field with increasing of $J/J_{\parallel}$, while the magnetic position of the first triple point moves towards a lower field. Therefore, increasing $J/J_{\parallel}$ leads to diminish boundary of the phase $\vert\text{nsns}\rangle_1$ with zero magnetization and to increase the boundary of phase $\vert\text{nsus}\rangle$ with $M/M_\text{s}=1/4$. Other phase boundaries will remarkably change, as well.
In the next part, we investigate the MCE of the Ising-Heisenberg ladder nearby the first triple point at which the coexistence of three ground states $\vert\text{nsns}\rangle_1$, $\vert\text{nsus}\rangle$ and $\vert\text{nunu}\rangle$ occurs, then we compare the result with that of obtained for the MCE of the Ising-Heisenberg model close to the quadruple point.

\begin{figure}
  \centering
  \resizebox{0.5\textwidth}{!}{
 \includegraphics[trim=20 10 40 20, clip]{ 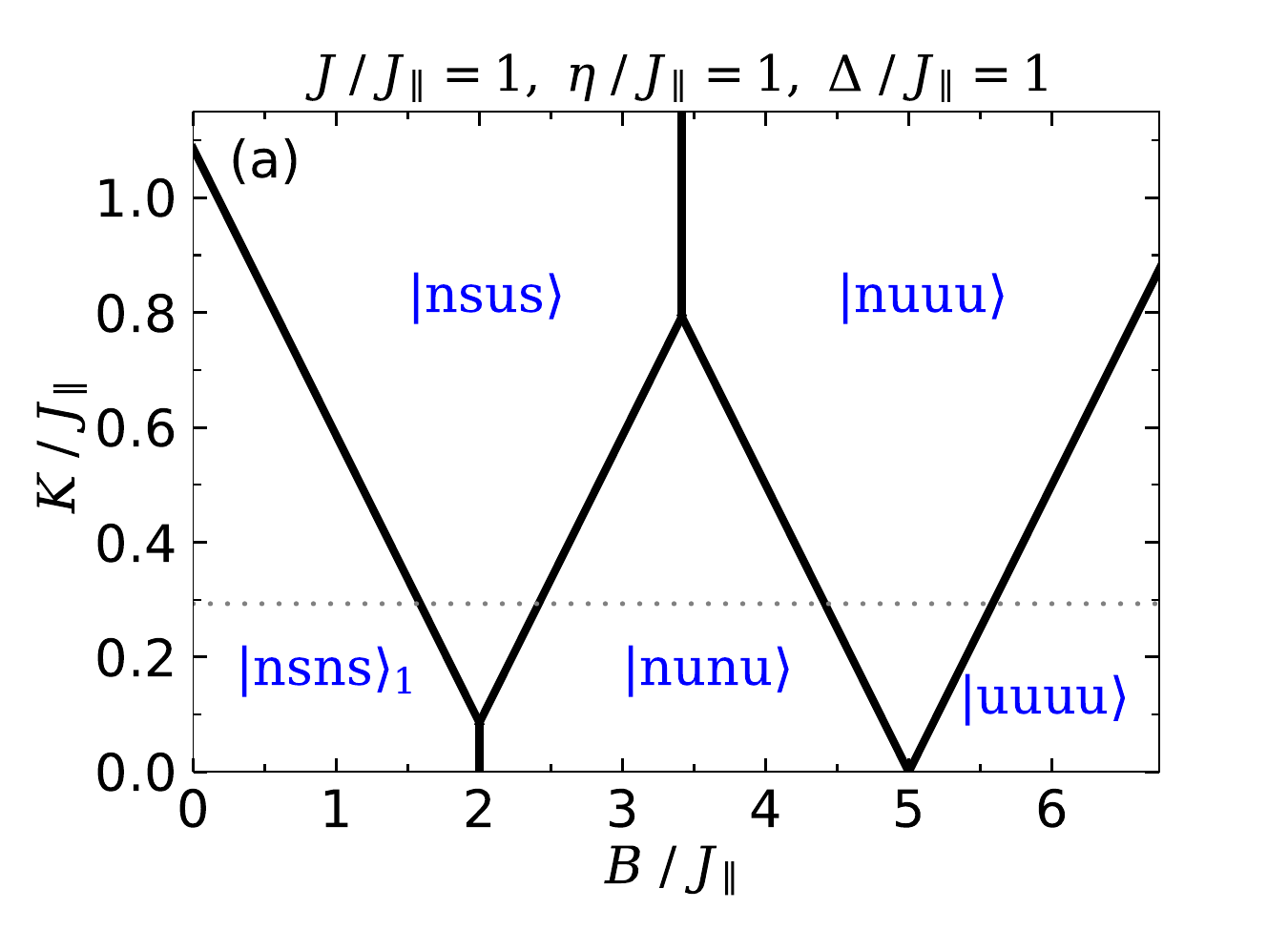 }
 \includegraphics[trim=50 10 40 20, clip]{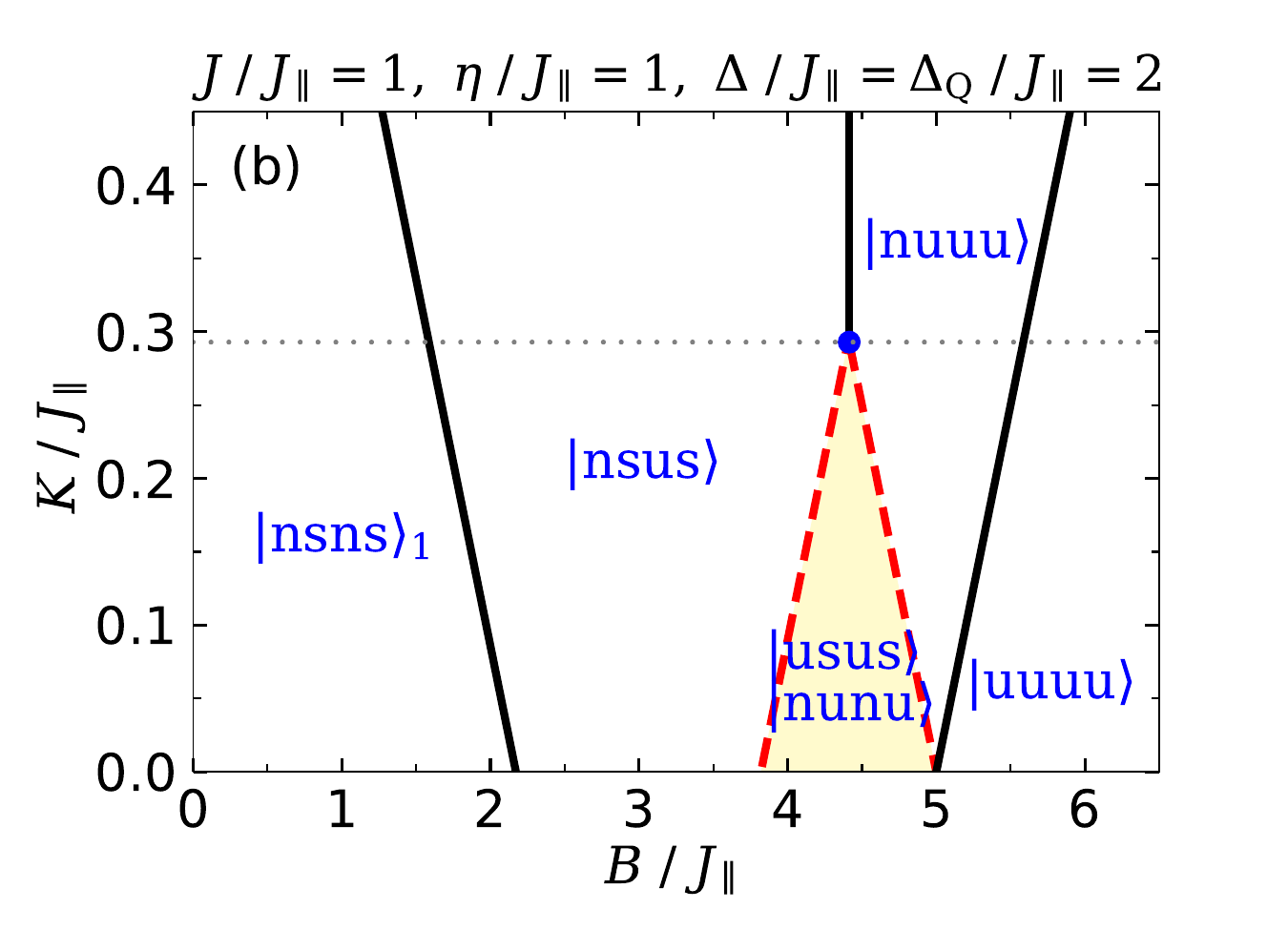}

}
\vspace{-0.65cm}       

\caption{Ground-state phase diagram of the spin-1/2 Ising-Heisenberg double sawtooth ladder in the $K/J_{\parallel}-B/J_{\parallel}$ plane for exchange interaction ratios
(a) $\eta/J_{\parallel}=J/J_{\parallel}=1$, $\Delta/J_{\parallel}=1$,
(b) $\eta/J_{\parallel}=J/J_{\parallel}=1$, $\Delta/J_{\parallel}=2$.
Red broken lines in panel (b) represent the same phase boundary for the ground states $\vert\text{nunu}\rangle$ and $\vert\text{usus}\rangle$, which become degenerate when Eq. (\ref{eq:Delta_Q}) holds for the interaction ratios $J_{\perp}/J_{\parallel}$ and $\eta/J_{\parallel}$. Solid dot indicates the position of the quadruple point according to Eqs. (\ref{eq:Delta_Q})-(\ref{eq:B_Q}) whose coordinates are $[K_\text{Q}/J_{\parallel}, B/J_{\parallel}]\approx[0.2929, 4.4142]$.
  The horizontal dotted lines indicate the value $K_\text{Q}/J_{\parallel}$ given by Eq. (\ref{eq:K_Q}). We note that  according to Eqs. (\ref{eq:K_Q}) and (\ref{eq:B_Q}) that hold for each point of the shaded area of Fig. \ref{fig:KQ}(b), having $[K,B]=[K_\text{Q},B_\text{Q}]$, thus the quadruple point is generated, if and only if, $\Delta/J_{\parallel}=\Delta_\text{Q}/J_{\parallel}=2.0$. }
\label{fig:GSPD_BK}
\end{figure}

In Fig. \ref{fig:GSPD_BK}, the ground-state phase diagram of the Ising-Heisenberg double sawtooth ladder is illustrated in the $K/J_{\parallel}-B/J_{\parallel}$ plane for two selected values of $\Delta/J_{\parallel} = \{1.0, 2.0\}$. In Fig. \ref{fig:GSPD_BK}(a) and we assume $\eta/J_{\parallel}=J/J_{\parallel}=1.0$ and plot the ground-state phase diagram of the Ising-Heisenberg model in the $K/J_{\parallel}-B/J_{\parallel}$ plane for $\Delta/J_{\parallel}=1.0$. Here, one sees five different ground states that are separated from each other by the corresponding boundary lines. There are two triple points at which three ground states coexist together.
  When $K/J_{\parallel}=0$, two discontinuous ground-state phase transition are observed in Fig. \ref{fig:GSPD_BK}(a). One transition is from $\vert\text{nsns} \rangle_1$ to $\vert\text{nunu}\rangle$ at critical magnetic field $B/J_{\parallel}=2.0$, and another one from $\vert\text{nunu}\rangle$ to fully polarized $\vert\text{uuuu}\rangle$ state at critical magnetic field $B/J_{\parallel}=5.0$.
The state $\vert\text{nuuu}\rangle$ with magnetization value $M/M_\text{s}=3/4$ could not be the ground state of the Ising-Heisenberg ladder for  $K/J_{\parallel}=0$. This scenario will also take place in the full Heisenberg model on the double sawtooth ladder when
$K/J_{\parallel}=0$ and $\Delta/J_{\parallel}=1.0$.
In Fig. \ref{fig:GSPD_BK}(b), we display the ground-state phase diagram of the Ising-Heisenberg model for $J/J_{\parallel}=\eta/J_{\parallel}=1.0$ and $\Delta/J_{\parallel}=\Delta_\text{Q}/J_{\parallel}=2.0$ (Eq. (\ref{eq:Delta_Q})).
Under these circumstances, the aforementioned quadruple point appears in the ground-state phase diagram of the model such that the two ground states $\vert\text{nunu}\rangle$ and $\vert\text{usus}\rangle$ become degenerate, whilst Eq. (\ref{eq:Delta_Q}) holds for the interaction ratios $J_{\perp}/J_{\parallel}$ and $\eta/J_{\parallel}$. A shaded region delimited by red dashed lines identifies the coexistence region of these two degenerate ground states with the identical magnetization ($M/M_\text{s}=1/2$). A solid point in Fig. \ref{fig:GSPD_BK}(b) indicates the coordinates of the quadruple point.

It is demonstrated in Fig. \ref{fig:KQ}(b) that the field coordinate of the quadruple point shifts towards lower magnetic field and lower four-spin Ising interaction values as the interaction ratio $J/J_{\parallel}$ decreases. For example, the coordinates of the quadruple point in Fig. \ref{fig:GSPD_BK}(b) are $[K_\text{Q}^\text{max}/J_{\parallel}, B_\text{Q}^\text{max}/J_{\parallel}]\approx[0.2929, 4.4142]$. Now, suppose $J/J_{\parallel}=0.5$, the coordinates of the quadruple point are $[K_\text{Q}/J_{\parallel}, B_\text{Q}/J_{\parallel}]\approx[0.19098, 3.62]$. One can see in Fig. \ref{fig:GSPD_BK}(b) that there are three different first-order phase transitions in the ground-state phase diagram of the Ising-Heisenberg ladder for $K/J_{\parallel}=0$ and $\Delta/J_{\parallel}=2.0$. Apparently, the state $\vert\text{nsus}\rangle$ with magnetization value $M/M_\text{s}=1/4$ is the ground state of the system for moderate magnetic fields. Under this condition, the state $\vert\text{nuuu}\rangle$ with the magnetization value $M/M_\text{s}=3/4$ still could not be the ground state of the Ising-Heisenberg double sawtooth ladder.

\begin{figure*}
  \centering
  \resizebox{1.\textwidth}{!}{
 \includegraphics[trim=30 30 60 20, clip]{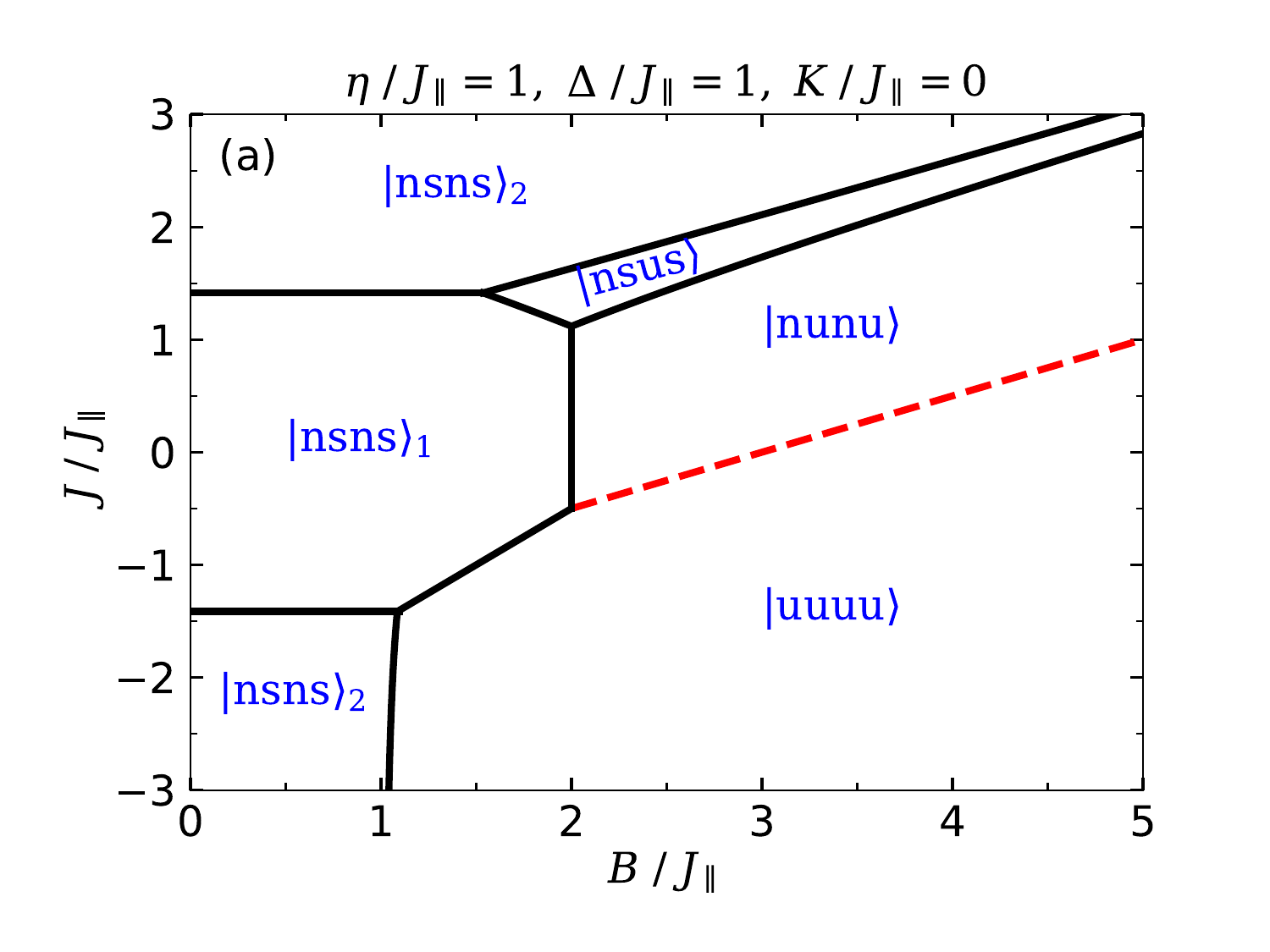}
 \includegraphics[trim=60 30 60 20, clip]{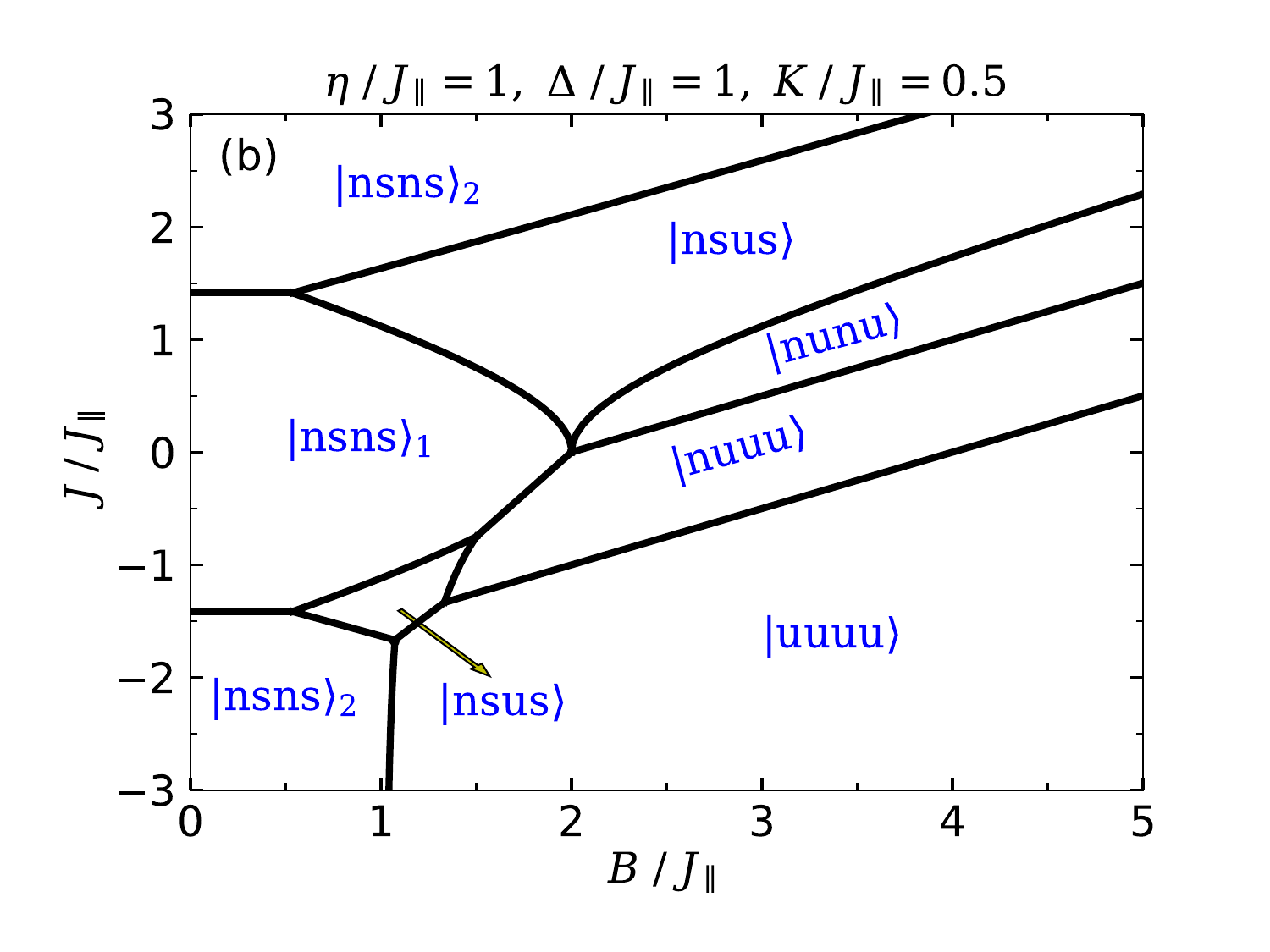}
 \includegraphics[trim=65 30 40 20, clip]{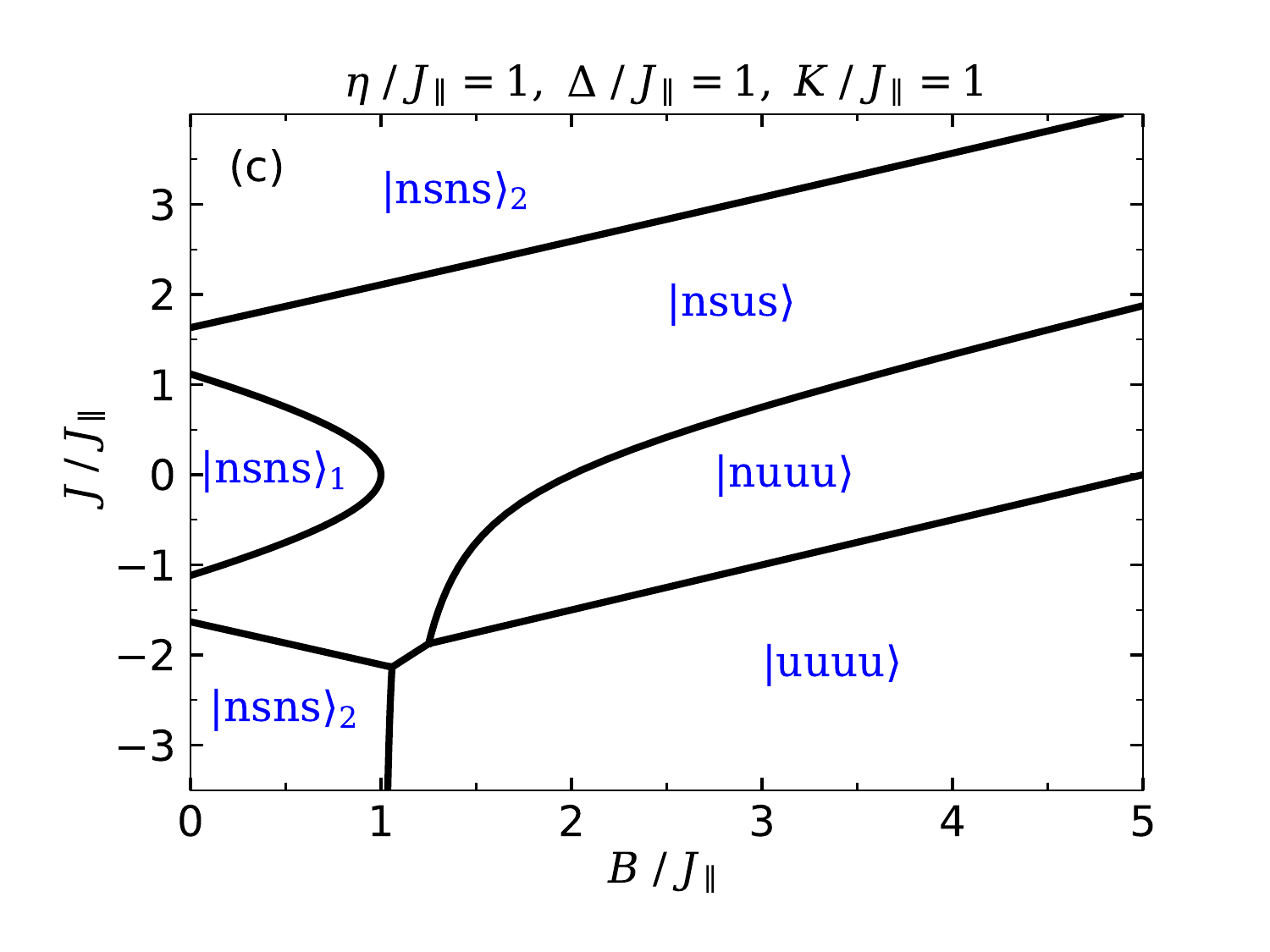}
}
\vspace{-0.65cm}       

\caption{Ground-state phase diagram of the spin-1/2 double sawtooth ladder in the $J/J_{\parallel}-B/J_{\parallel}$ plane for fixed values of $\eta/J_{\parallel}=J/J_{\parallel}=\Delta/J_{\parallel}=1$ and (a) $K/J_{\parallel}=0$, (b) $K/J_{\parallel}=0.5$ and (c) $K/J_{\parallel}=1$. Red dashed line in panel (a) describe the phase boundary between ground states $\vert\text{nunu}\rangle$ and $\vert\text{uuuu}\rangle$. For each point on this line, two ground states $\vert\text{nunu}\rangle$ and $\vert\text{nuuu}\rangle$ are degenerate. In panel (b), all multi-coexistence points are triple points.}
\label{fig:GSPD_JB}
\end{figure*}

The ground-state phase diagram of the spin-1/2 Ising-Heisenberg model on the double sawtooth ladder in the $J/J_{\parallel}-B/J_{\parallel}$ plane is displayed in Fig. \ref{fig:GSPD_JB} for the parameter set $\eta/J_{\parallel}=\Delta/J_{\parallel}=1$ supposing the isotropic interaction betwen the Heisenberg spins. Figure \ref{fig:GSPD_JB}(a) represents the ground-state phase diagram of the Ising-Heisenberg ladder for $K/J_{\parallel}=0$. Except the ground states already emergent in Fig. \ref{fig:GSPD_BK} one additionally detects  in the parameter regions $J/J_{\parallel}<-1.4142$ and $J/J_{\parallel}>1.4142$ the antiferromagnetic  ground state $\vert\text{nsns}\rangle_2$. Two ground states $\vert\text{nunu}\rangle$ and $\vert\text{nuuu}\rangle$ become degenerate on the red broken line with the condition $J/J_{\parallel}=\dfrac{1}{2}B/J_{\parallel}-\dfrac{3}{2}$. This line separates two ground states $\vert\text{nunu}\rangle$ and $\vert\text{uuuu}\rangle$. By inspecting Fig. \ref{fig:GSPD_BK}(b), one can find out that even a small but nonzero value of the four-spin Ising interaction ratio $K/J_{\parallel}$ considerably changes a stability regions of the ground states such that new phase boundaries are formed. For example, the phase boundary of the ground state $\vert\text{nuuu}\rangle$ gradually broadens with increasing of the ratio $K/J_{\parallel}$  , while the phase boundary of the ground state $\vert\text{nunu}\rangle$ diminishes until it ultimately disappears (see Fig. \ref{fig:GSPD_BK}(c)). Besides, the four-spin exchange interaction gives rise to the ground state $\vert\text{nsus}\rangle$ emergent in a ferromagnetic regime of the interaction ratio $J/J_{\parallel}<0$. It could be concluded that the stability region of the ground state $\vert\text{nsus}\rangle$ and $\vert\text{nuuu}\rangle$ increases for $\eta/J_{\parallel}=\Delta/J_{\parallel}=1$ with increasing of $K/J_{\parallel}$, while the stability region of the ground state $\vert\text{nunu}\rangle$ decreases until it completely disappears.

\subsection{Adiabatic (de)magnetization process of the spin-1/2   Ising-Heisenberg ladder}\label{cooling}
Recently, many authors have widely reported that various frustrated quantum spin systems exhibit an enhanced magnetocaloric effects (MCE) during the adiabatic demagnetization process, which may be of principle importance for the low-temperature magnetic refrigeration. Hence, let us also investigate the adiabatic demagnetization process of the spin-1/2  Ising-Heisenberg double sawtooth ladder under the particular adiabatic conditions. In the following we will study the isentropes (levels of constant entropy)  in the $B/J_{\parallel}-T/J_{\parallel}$ plane and the magnetic Gr{\"u}neisen parameter times temperature, $T\Gamma_B$ ,as a function of the magnetic field close to the triple and quadruple points, which have been comprehensively described in the previous section. The effects of the exchange anisotropy $\Delta/J_{\parallel}$ on the entropy and the magnetic cooling rate are investigated as well.

\begin{figure*}
\begin{center}
\resizebox{1\textwidth}{!}{%
\includegraphics[trim=40 70 50 50, clip]{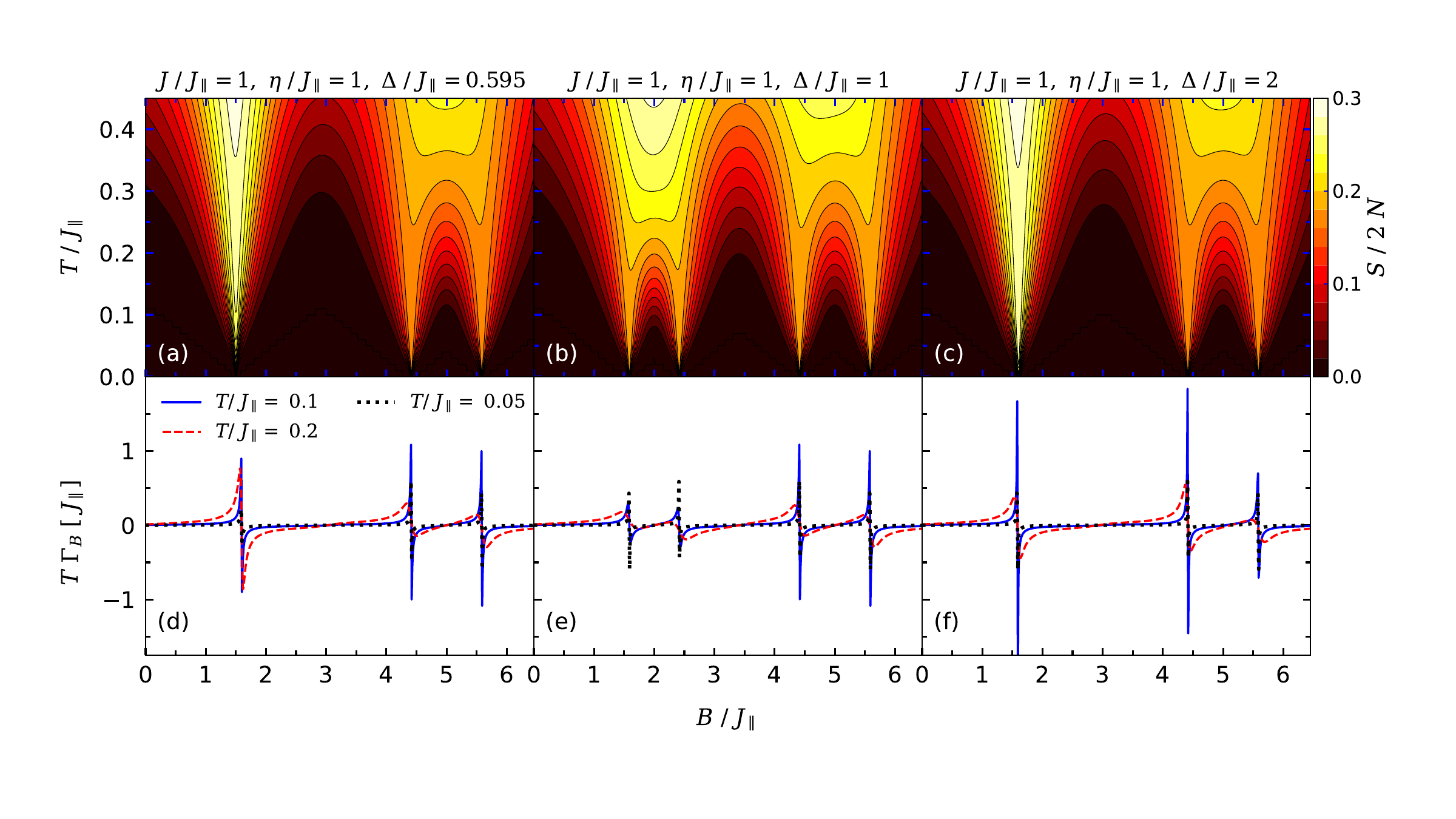}
}
\vspace{-0.65cm}
\caption{(a)-(c) Contour plots of the isentropes of the spin-1/2 Ising-Heisenberg double sawtooth ladder in the temperature $T/J_{\parallel}$versus the magnetic field $B/J_{\parallel}$ plane for the fixed value of $J/{J}_{\parallel}=\eta/J_{\parallel}=1$. According to Eq. (\ref{eq:K_Q}), the four-spin exchange term is taken as $K=K_\text{Q}^\text{max}\approx 0.2929$ in all panels. In panel (a) the corresponding coordinates of the first triple point is taken as $\Delta/J_{\parallel}=0.595$ (see Fig. \ref{fig:GSPD_BDelta}(b)). In panel (b) we consider the isotropic coupling constant $\Delta/J_{\parallel}=1.0$. In panel (c) according to the coordinates of the quadruple point in Fig. \ref{fig:GSPD_BDelta}(b), we consider the exchange anisotropy as $\Delta/J_{\parallel}=2.0$.
 By inspecting Fig. \ref{fig:GSPD_BDelta}(b), each temperature drop occurs nearby the first-order ground-state phase transition.
 (d)-(f) The magnetic Gr{\"u}neisen parameter multiplied by temperature $T\Gamma_B$ as a function of magnetic field $B/J_{\parallel}$ for the same set of parameters to panels (a)-(c), respectively. Three different temperatures $T/J_{\parallel}=\{0.05, 0.1, 0.2\}$ are considered.}
\label{fig:STBH}
\end{center}
\end{figure*}

Figures \ref{fig:STBH}(a)-(c) show the entropy of spin-1/2 Ising-Heisenberg double sawtooth ladder in the $B/J_{\parallel}-T/J_{\parallel}$ plane by assuming the following parameter values $J/J_{\parallel}=\eta/J_{\parallel}=1$ and three different exchange anisotropies $\Delta/J_{\parallel}=\{0.595, 1.0, 2.0\}$. It is worthwhile to recall that Eq. (\ref{eq:K_Q}) holds for the particular case $K=K_\text{Q}^\text{max}\approx 0.2929$ (see horizontal dotted lines in Figs. \ref{fig:GSPD_BK}(a)-(c)). By comparing Fig. \ref{fig:STBH}(a) with the ground-state phase diagram depicted in Fig. \ref{fig:GSPD_BDelta}(b), one encounters that the sizable MCE in a low-temperature regime with relatively low value of the entropy $S/4 N$ coincides with first-order ground-state phase transitions. It can be understood that if the magnetic field decreases adiabatically from $B/J_{\parallel}=2.0$ towards the triple point $B/J_{\parallel}=1.5858$ the temperature dramatically drops down.

By increasing of the exchange anisotropy $\Delta/J_{\parallel}$ the value of entropy significantly changes nearby the critical fields, at which temperature falls down (see Figs.  \ref{fig:STBH}(b) and (c)). Evidently, the alterations of entropy $S/4 N$ is sharper close to the triple and quadruple points where the temperature drops down more rapidly. Generally, we observe three enhanced regions of MCE accompanied with a relatively fast cooling of the model at two selected   coordinates associated to the triple and quadruple points of Fig. \ref{fig:GSPD_BDelta}(b) (see Figs. \ref{fig:STBH}(a) and (c)). In Fig. \ref{fig:STBH}(a), the first region of enhanced MCE is due to existence of the triple point at which three ground states $\vert\text{nsns} \rangle_1$, $\vert\text{nsus}\rangle$ and $\vert\text{nunu}\rangle$ coexist together. The second region of enhanced MCE is caused by the field-induced ground-state phase transition from $\vert\text{nunu}\rangle$ to $\vert\text{nuuu}\rangle$, while the third one appears nearby the field-driven ground-state phase transition from $\vert\text{nuuu}\rangle$ to $\vert\text{uuuu}\rangle$. Analogously, it can be observed that the modulation of the interaction parameters $\Delta/J_{\parallel}$ and $K/J_{\parallel}$ results in a change of the position of critical points at which the enhanced MCE occurs (see Figs. \ref{fig:STBH} (b) and (d)).

Now, let us discuss the effects of the exchange anisotropy and the cyclic four-spin Ising coupling on the magnetic Gr{\" u}neisen parameter. To this end, we plot in Figs. \ref{fig:STBH}(d)-(f) the magnetic Gr{\" u}neisen parameter times temperature, $T\Gamma_B$, against the magnetic field at three different temperatures: $T/J_{\parallel}=\{0.05, 0.1, 0.2\}$ for the same set of the interaction parameters as used in Figs. \ref{fig:STBH}(a)-(c), respectively. It follows from Fig. \ref{fig:STBH}(d) that the product $T\Gamma_B$ changes its sign nearby  the triple point emergent at $B/J_{\parallel}\approx 1.5858$, which is closely related to an accumulation of the isentropy lines due to the coexistence of three different ground states.
Setting up all parameters such that four ground states $\vert \text{nsus} \rangle$, $\vert\text{nunu}\rangle$, $\vert\text{usus} \rangle$ and $\vert\text{nuuu}\rangle$ coexist together at the quadruple point with coordinates $[\Delta/J_{\parallel}, K_\text{Q}^\text{max}/J_{\parallel}, B_\text{Q}^\text{max}/J_{\parallel}]=[2.0, 0.2929, 4.4142]$ results in observing a fast cooling/heating of the Ising-Heisenberg system with even larger value of the Gr{\" u}neisen parameter.
Therefore, the magnetic behavior of Gr{\" u}neisen parameter corroborates that the MCE is considerably enhanced during the adiabatic demagnetization process if the interaction parameters  $[K/J_{\parallel},\Delta/J_{\parallel}]$ are tuned close enough to the quadruple point. From the exact results obtained for the entropy and cooling rate of the Ising-Heisenberg double sawtooth ladder shown in Figs. \ref{fig:STBH}(c) and (f) one can deduce that the MCE is remarkably enhanced in a close vicinity of the quadruple point.
It could be indeed concluded that the spin-1/2 Ising-Heisenberg double sawtooth ladder exhibits the enhanced MCE compared to other ranges of $K/J_{\parallel}\neq K_\text{Q}/J_{\parallel}$ whenever Eqs. (\ref{eq:K_Q}) and (\ref{eq:B_Q}) hold for each point on the curves plotted in Fig. \ref{fig:KQ}(b).

Another result gained from our examinations is the particular response of the magnetic Gr{\"u}neisen parameter with respect to temperature variations. We display in Figs. \ref{fig:STBH}(d)-(f) the product $T\Gamma_B$ as a function of the magnetic field at three different temperatures as well. As usual, increasing of the temperature suppresses in general the enhanced MCE occurred at critical magnetic fields. The interesting point to declare is that the observed MCE at triple and quadruple points shows an unconventional resistance against the rising temperature. As a matted of fact, the enhanced MCE is sizable close to the triple and quadruple points even at higher temperatures when $\Delta/J_{\parallel}=2.0$m $K/J_{\parallel}=K_\text{Q}/J_{\parallel}$ and the magnetic field is tuned sufficiently close to the critical point $B_\text{Q}/J_{\parallel}$.

\section{Magnetization process of the spin-1/2 Heisenberg double sawtooth ladder}\label{fullHeisenberg}
In this section, we proceed to study of the zero-temperature magnetization process of the full quantum spin-1/2 Heisenberg model on the double sawtooth ladder with four-spin Ising coupling in the presence of the external magnetic field. Unlike the previous case, the magnetic ground states of the fully quantum spin-1/2 Heisenberg double sawtooth ladder cannot be rigorously extracted through exact analytical methods and hence, the exact diagonalization (ED) exploiting the Lanczos algorithm and density matrix renormalization group (DMRG) methods are used as two numerical techniques to solve the Hamiltonian of the spin-1/2 Heisenberg double sawtooth ladder given by:

\begin{eqnarray}\label{Heis_hamiltonian}
H =
&& \sum\limits_{j=1}^N\Big[\left({\boldsymbol \sigma}_{1,j}\cdot{\boldsymbol \sigma}_{2,j}\right)_{\eta,\Delta}+
\nonumber \\
&& \sum\limits_{a=1,2}\big({J}_{\parallel}{\boldsymbol \sigma}_{a,j}\cdot{\boldsymbol \sigma}_{a,j+1}+ J {\boldsymbol \sigma}_{a,j}\cdot({\boldsymbol \sigma}_{a,j-1}+{\boldsymbol \sigma}_{a,j+1})\big)
\nonumber \\
&& +{J}_{\perp}{\boldsymbol \sigma}_{1,j}\cdot{\boldsymbol \sigma}_{2,j}+K{\sigma}_{1,j}^z{\sigma}_{1,j+1}^z{\sigma}_{2,j}^z{\sigma}_{2,j+1}^z\nonumber\\
&& -B\sum\limits_{a=1,2}\sigma_{a,j}^z\Big].
\end{eqnarray}

\begin{figure*}
\centering
\resizebox{1\textwidth}{!}{
 \includegraphics[trim=65 60 100 70, clip]{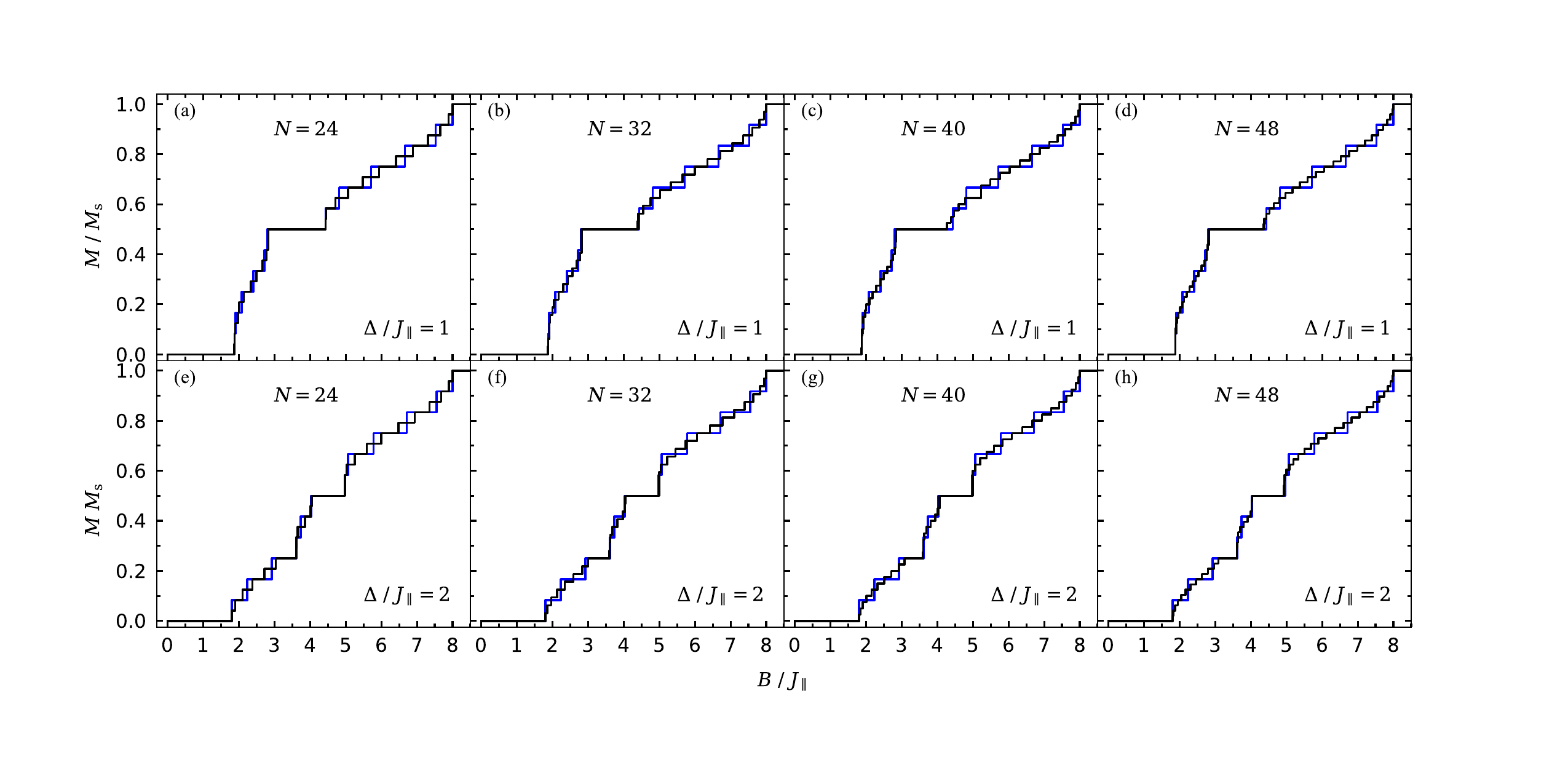}
}
\vspace{-0.65cm}       

\caption{Full ED (blue lines) and DMRG (black lines) results for zero-temperature magnetization curves of the spin-1/2 Heisenberg double sawtooth ladder. The interaction ratios $\eta/J_{\parallel}=J/J_{\parallel}=1$ and $K/J_{\parallel}=0$ are considered in all panels. The ED data were obtained for $N-12$, whereas the DMRG data were obtained for the following set of parameters:
(a) $\Delta/J_{\parallel}=1$, $N=24$,
(b) $\Delta/J_{\parallel}=1$, $N=32$,
(c) $\Delta/J_{\parallel}=1$, $N=40$,
(d) $\Delta/J_{\parallel}=1$, $N=48$,
(e) $\Delta/J_{\parallel}=2$, $N=24$,
(f) $\Delta/J_{\parallel}=2$, $N=32$,
(g) $\Delta/J_{\parallel}=2$, $N=40$,
(h) $\Delta/J_{\parallel}=2$, $N=48$.
}
\label{fig:Mag_dmrgLancz}
\end{figure*}

\begin{figure*}
\begin{center}
\resizebox{1\textwidth}{!}{%
\includegraphics[trim=20 00 50 50, clip]{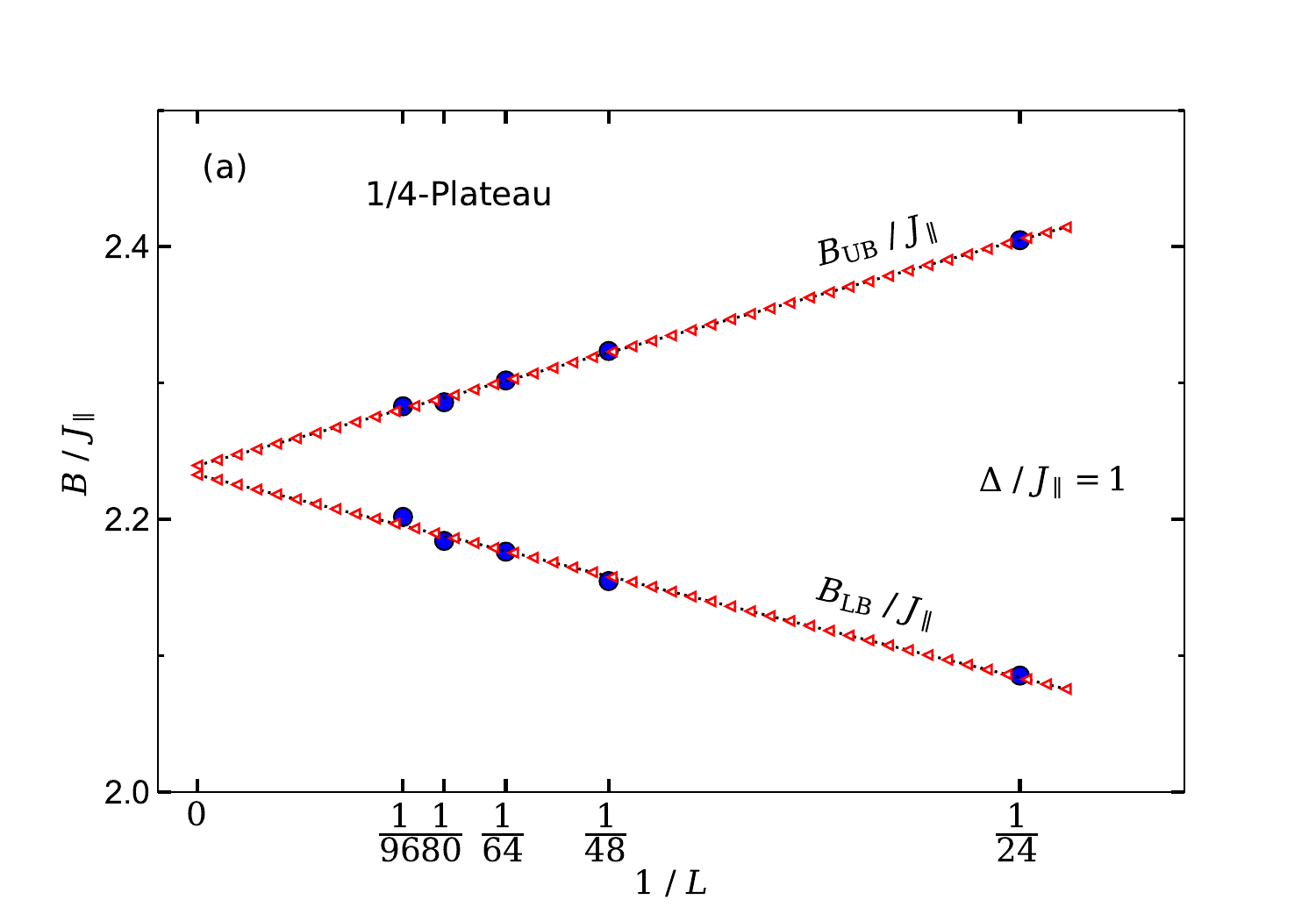}
\includegraphics[trim=50 0 50 50, clip]{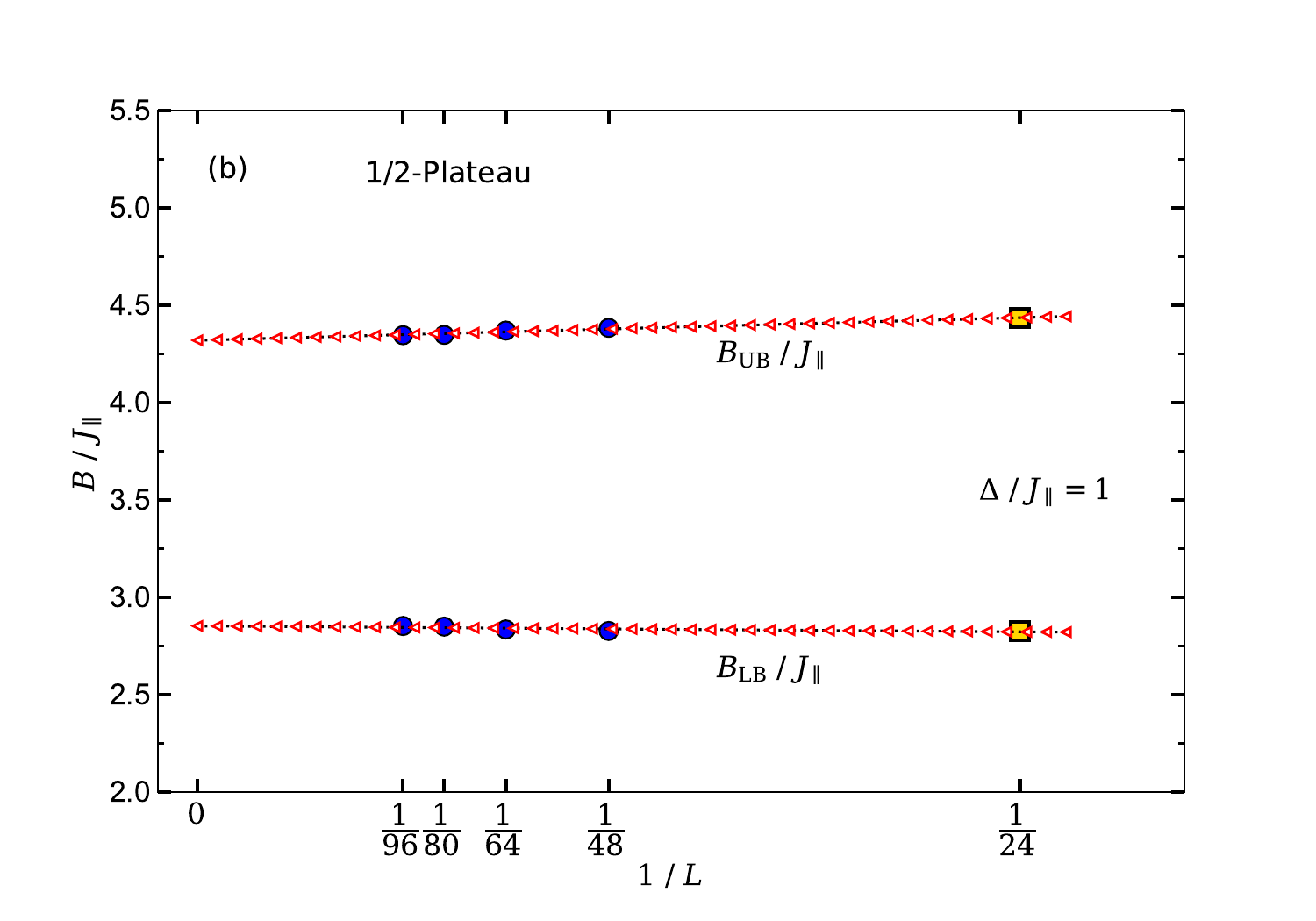}
\includegraphics[trim=50 0 20 50, clip]{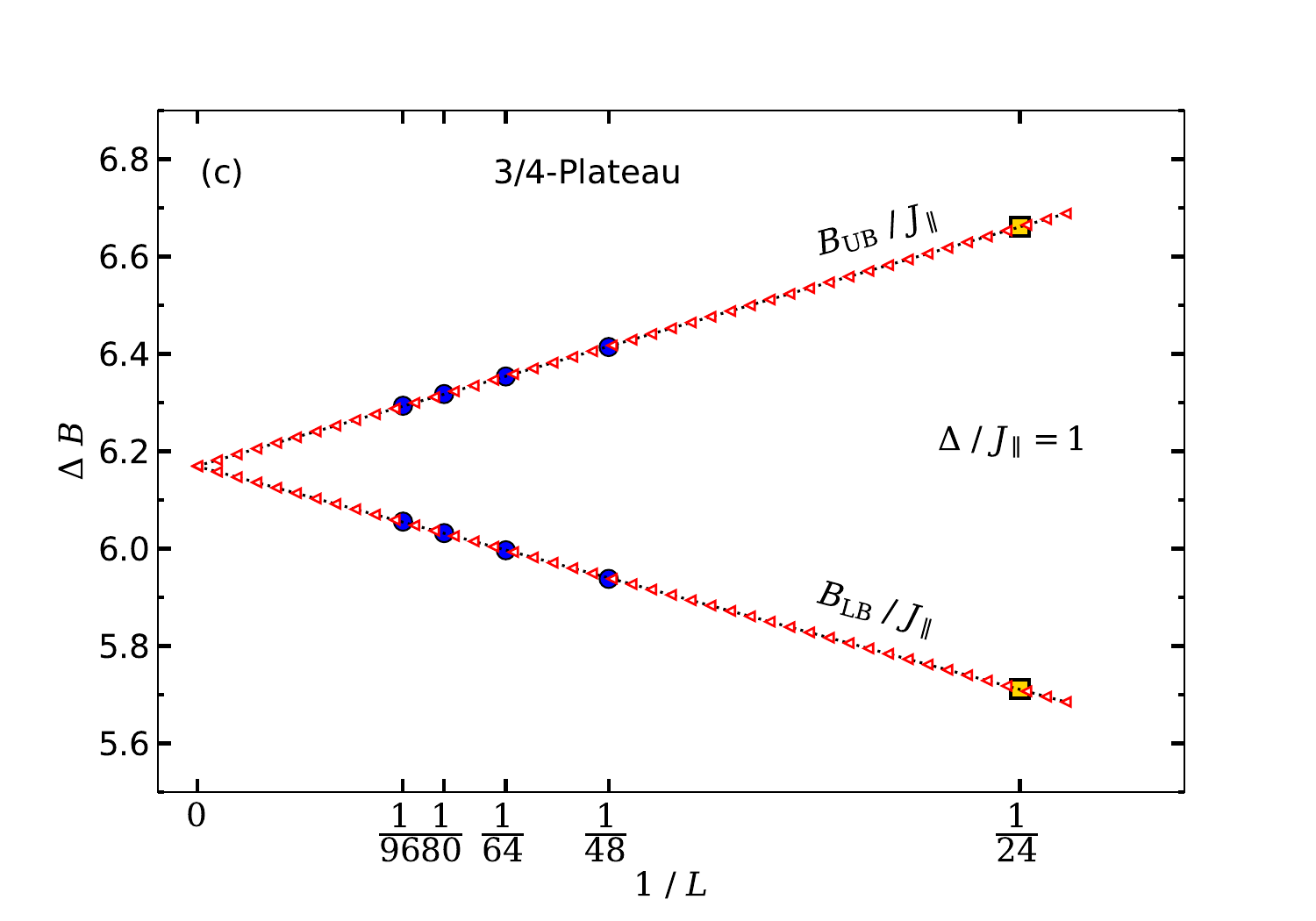}
}
\resizebox{1\textwidth}{!}{%
\includegraphics[trim=20 0 50 60, clip]{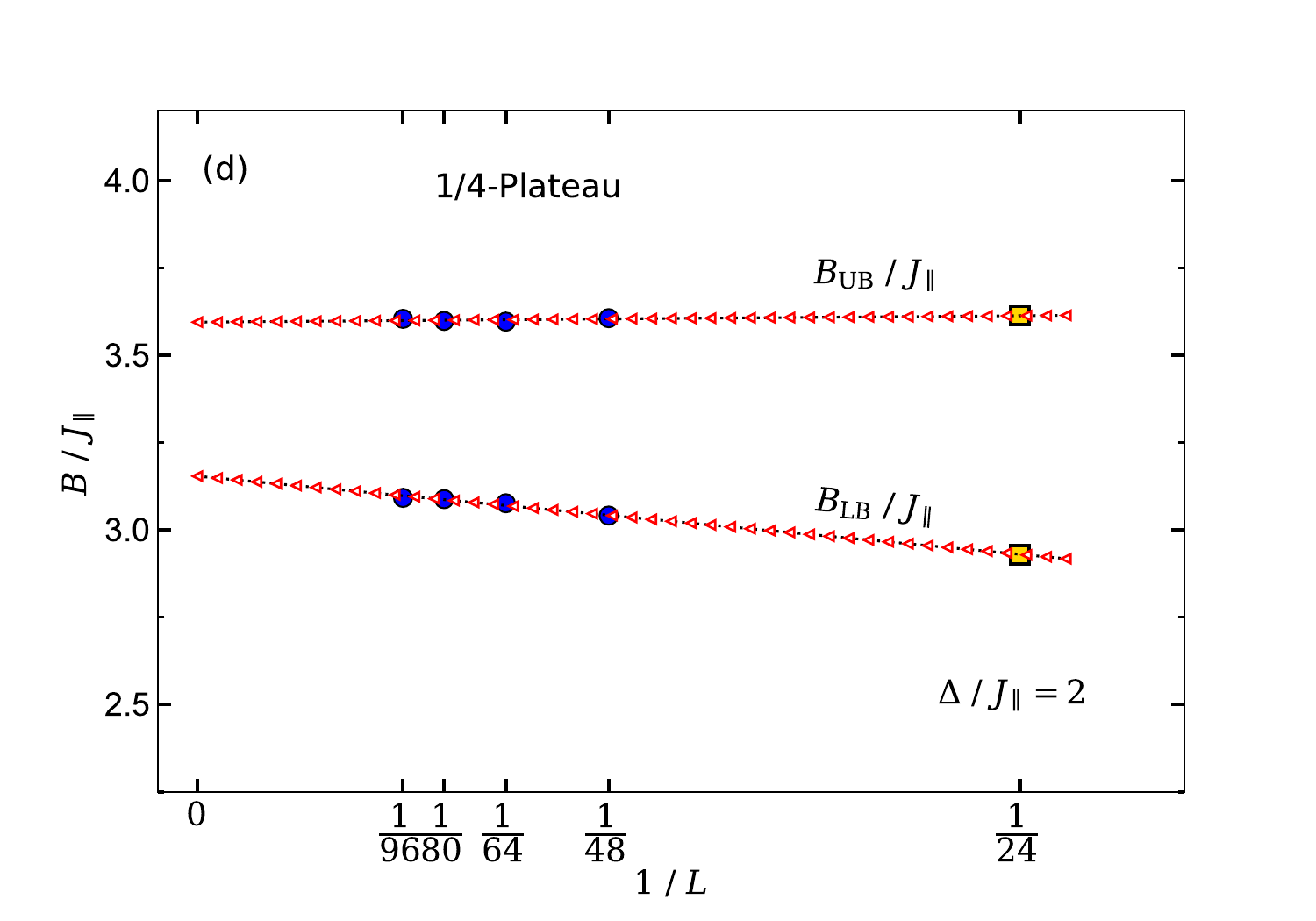}
\includegraphics[trim=50 0 50 60, clip]{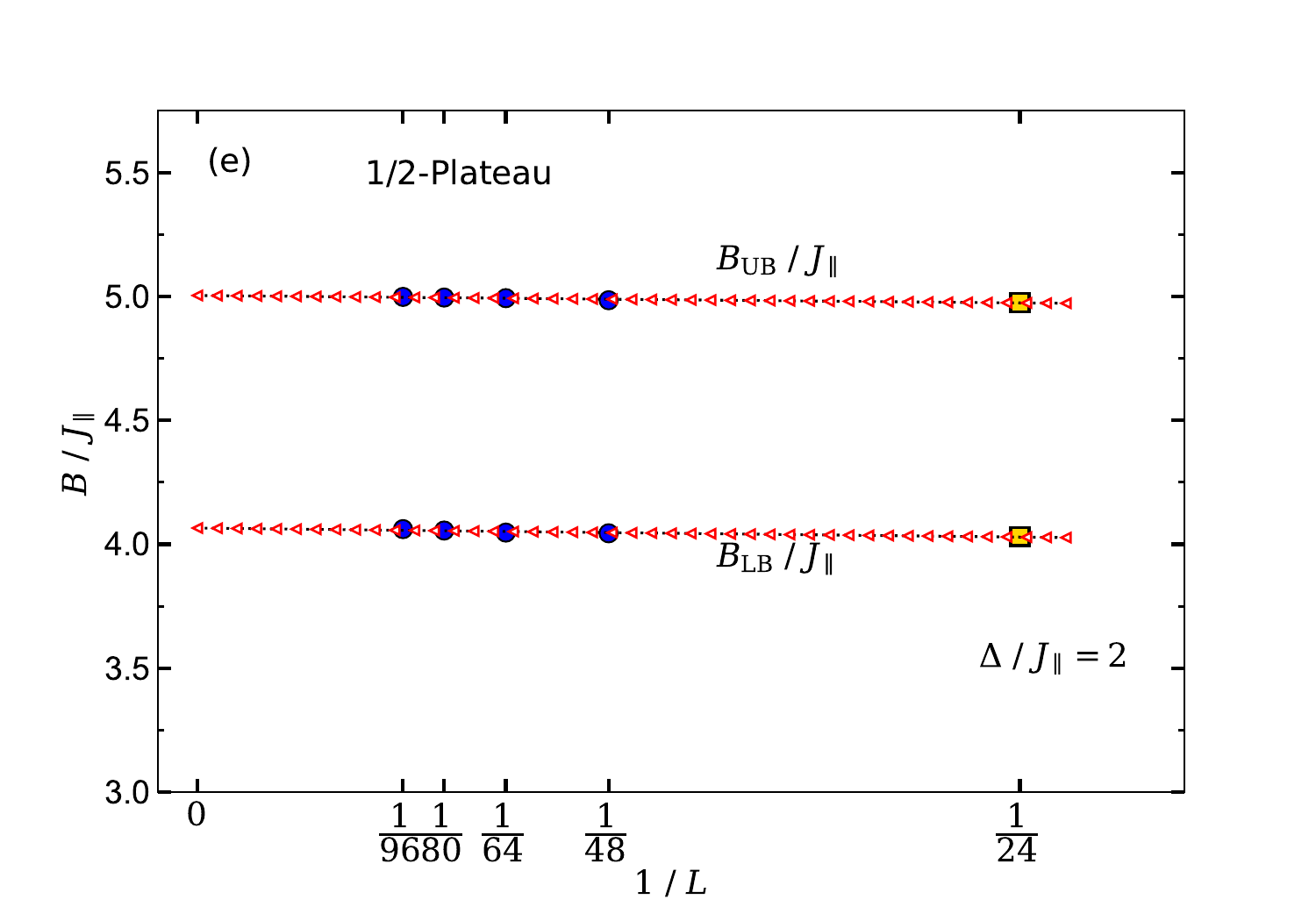}
\includegraphics[trim=50 0 20 60, clip]{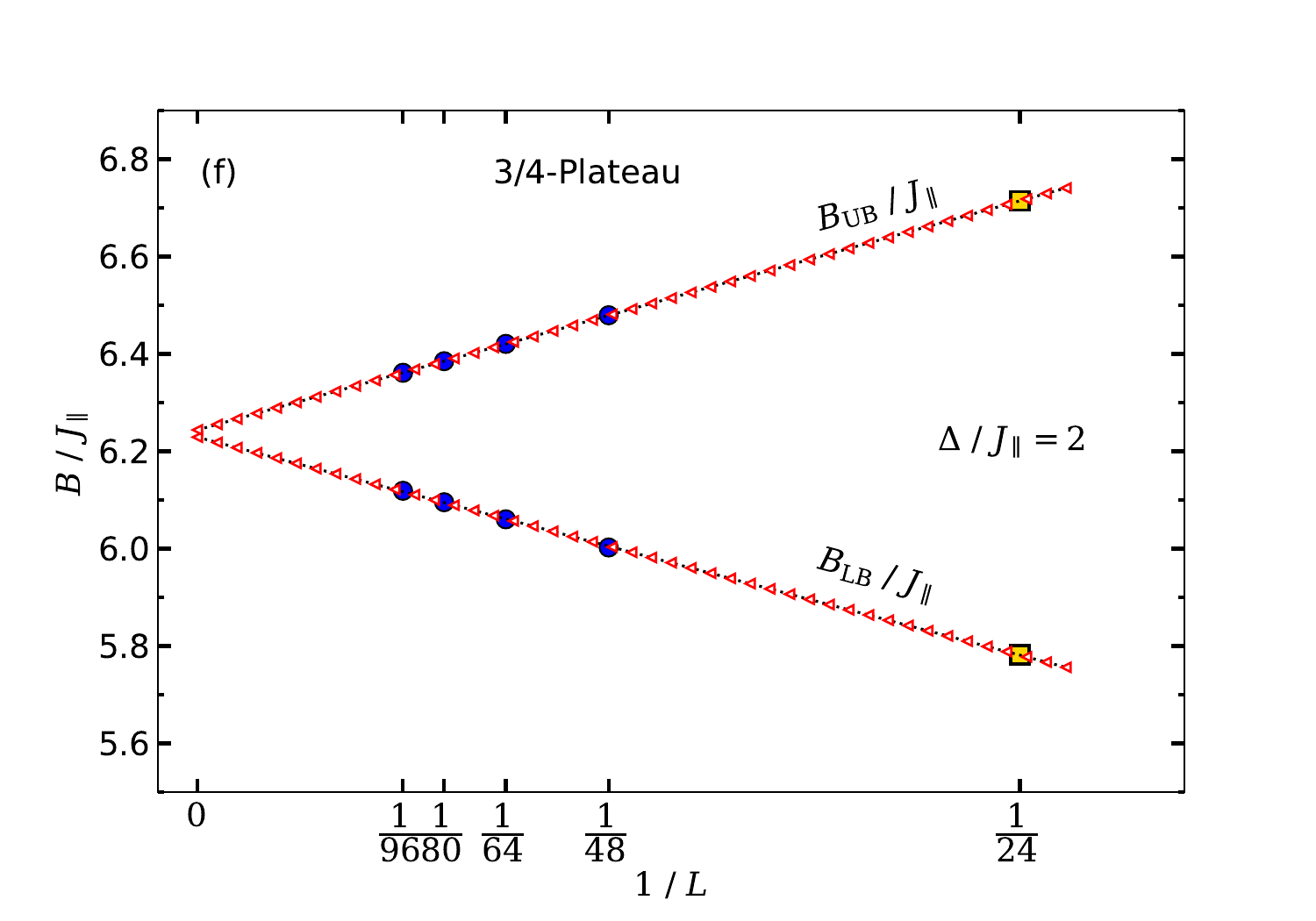}

}
\vspace{-0.65cm}
\caption{The magnetic fields $B_\text{LB}/J_{\parallel}$ and $B_\text{UB}/J_{\parallel}$ corresponding to lower and upper bounds of the intermediate magnetization plateaus (DMRG data - circles, ED data - squares) together with the relevant extrapolation to the thermodynamic limit $N\rightarrow \infty$	 by considering: (a) one-quarter plateau for $\Delta/J_{\parallel}=1$, (b) one-half plateau for $\Delta/J_{\parallel}=1$, (c) three-quarter plateau for $\Delta/J_{\parallel}=1$, (d) one-quarter plateau for $\Delta/J_{\parallel}=2$, (e) one-half plateau for $\Delta/J_{\parallel}=2$, (f) three-quarter plateau for $\Delta/J_{\parallel}=2$.}
\label{fig:dB_N}
\end{center}
\end{figure*}

By using ED method, we have exactly computed the lowest-energy eigenstates of the spin-1/2 Heisenberg model on the double sawtooth ladder
for relatively small finite-size systems up to $N=12$ (i.e. the total number of spins 24). These exact numerical results will be subsequently confronted with the DMRG data calculated for the spin-1/2 Heisenberg double sawtooth ladder for larger finite-size systems with $N=24, 32, 40, 48$ (i.e. the total number of spins 48, 64, 80, 96). Firstly, let us consider absence of the four-spin coupling (i.e. $K/J_{\parallel}=0$) in order to obtain zero-temperature magnetization curves of the full spin-1/2 Heisenberg double sawtooth ladder for two different values of the exchange anisotropy $\Delta/J_{\parallel}=1$ and $\Delta/J_{\parallel}=2$ (see Fig. \ref{fig:Mag_dmrgLancz}). Figure \ref{fig:Mag_dmrgLancz} confronts the numerical results for zero-temperature magnetization curves as obtained from the ED and DMRG methods for the fixed values of $\eta/J_{\parallel}=\Delta/J_{\parallel}=J/J_{\parallel}=1$ and $K/J_{\parallel}=0$. It is evident from Fig. \ref{fig:Mag_dmrgLancz}(a)-(d) that the magnetization curves definitely exhibit zero and one-half magnetization plateaus when the total magnetization is normalized with respect to the saturation magnetization, while presence of other plateaus is questionable. To shed light on this issue, the width of the intermediate magnetization plateaus at one-quarter, one-half and three-quarter of the saturation magnetization is plotted in Figs. \ref{fig:dB_N}(a)-(c) for $\Delta/J_{\parallel}=1$. In Fig. \ref{fig:dB_N}(a), one observes that the intermediate one-quarter magnetization plateau gradually shrinks with increasing of the system size and it entirely disappears in the thermodynamic limit $N\rightarrow \infty$. Contrary to this, it is evident from Fig. \ref{fig:dB_N}(b) that the width of intermediate one-half magnetization plateau firmly survives in the thermodynamic limit $N\rightarrow \infty$. Last but not least, it follows from Fig. \ref{fig:dB_N}(c) that the intermeditae three-quarter magnetization plateau gradually disappears in the thermodynamic limit and hence, it does not represent the true magnetization plateau for $\Delta/J_{\parallel}=1$. It could be concluded that the intermediate zero- and one-half magnetization plateaus of the spin-1/2 Heisenberg double sawtooth ladder survive in the thermodynamic limit, whereby the analogous intermediate plateaus of the spin-1/2 Ising-Heisenberg double sawtooth ladder correspond to the ground states $\vert\text{nsns}\rangle_1$ and $\vert\text{nunu}\rangle$. The DMRG simulations of the spin-1/2 Heisenberg double sawtooth ladder thus decisively verify presence or absence of given intermediate magnetization plateaus in the thermodynamic limit.

The ED and DMRG results for zero-temperature magnetization curves of the spin-1/2 Heisenberg double sawtooth ladder are plotted in Figs. \ref{fig:Mag_dmrgLancz}(e)-(h) for the particular case $\Delta/J_{\parallel}=2$. It could be anticipated that the magnetization curves of spin-1/2 Heisenberg double sawtooth ladder may display intermediate magnetization plateaus at zero, one-quarter and one-half of the saturation magnetization even for very large number of spins. Based on the DMRG calculations, it can be conjectured that two intermediate magnetization plateaus at one-quarter and one-half of the saturation magnetization are indeed the actual magnetization plateaus persisting in the thermodynamic limit. To bear evidence of this statement we plot in Figs. \ref{fig:dB_N}(d)-(f) the width of the intermediate magnetization plateaus against the inverse value of the total number of spins including results extrapolated to the thermodynamic limit. For two aforementioned intermediate magnetization plateaus, the lower and upper edges of the intermediate magnetization plateaus do not converge in the thermodynamic limit to the same asymptotic value, which means that they both represent true intermediate magnetization plateaus of the full spin-1/2 Heisenberg double sawtooth ladder with $\Delta/J_{\parallel}=2$. Similarly as in the previous case, the intermediate plateau at a three-quarter of the saturation magnetization may not be a true plateau, because the associated energy gap is closed when extrapolated to the thermodynamic limit.

\begin{figure*}
\centering
\resizebox{0.4\textwidth}{!}{
\includegraphics[trim=100 80 40 50, clip]{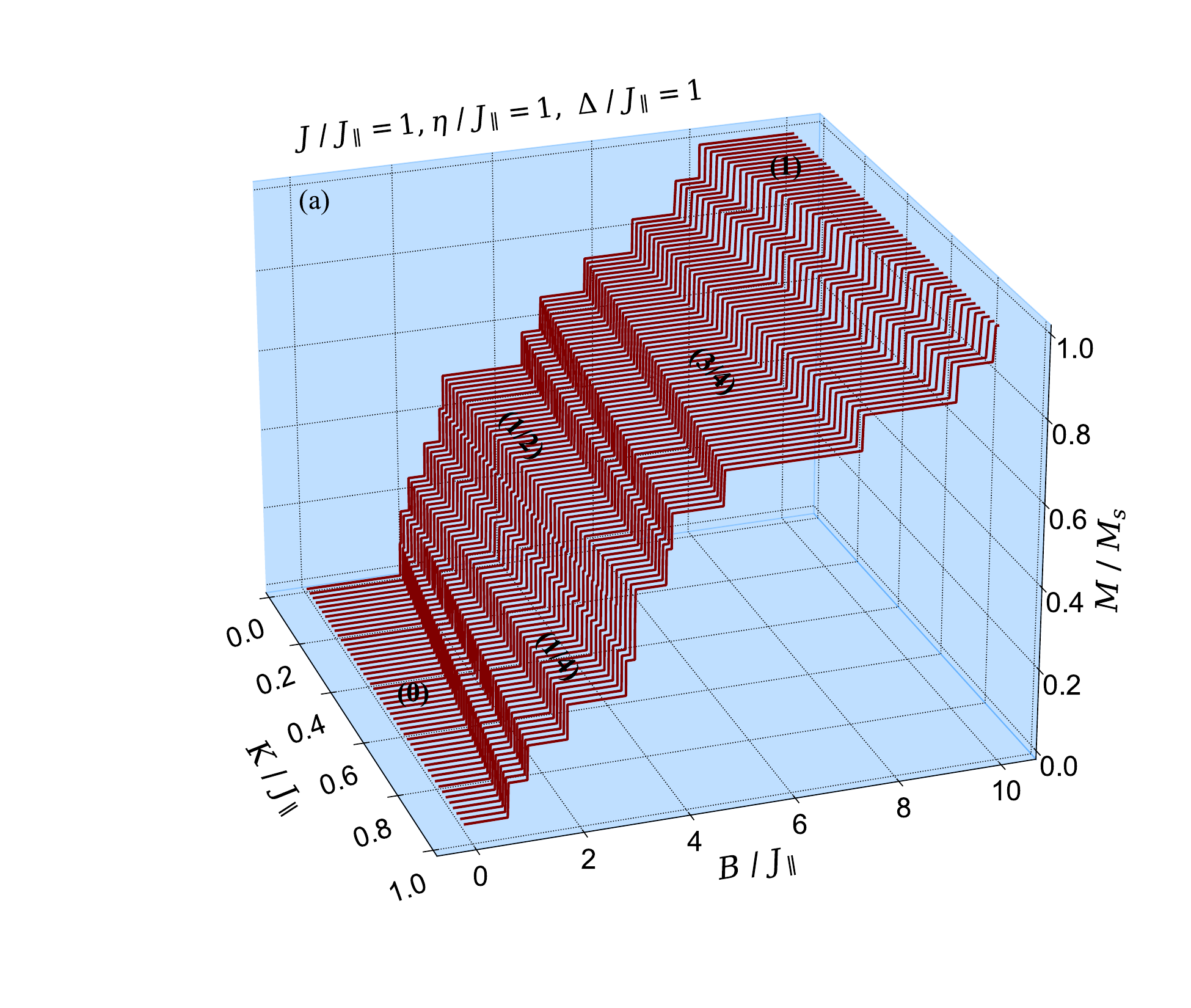}
 }
\resizebox{0.4\textwidth}{!}{
\includegraphics[trim=100 80 40 50, clip]{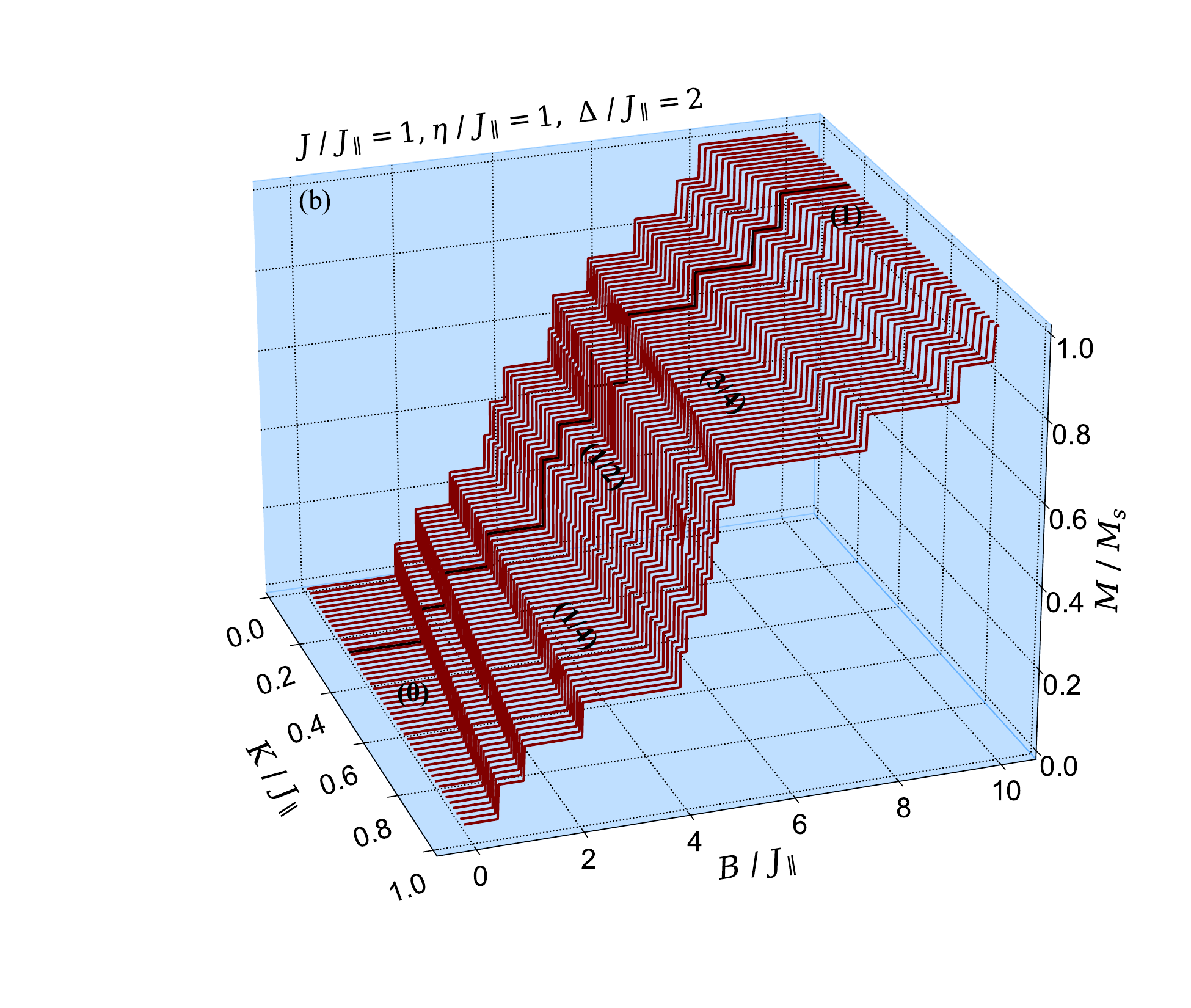}
}
\vspace{-0.25cm}       
\caption{3D plot of the zero-temperature magnetization of the spin-1/2 quantum Heisenberg double sawtooth ladder as a function of the magnetic field $B/J_{\parallel}$ and four-spin exchange interaction $K/J_{\parallel}$ for fixed value of the interaction parameters $\eta/J_{\parallel}=J/J_{\parallel}=1$ and two selected values of the exchange anisotropy: (a) $\Delta/J_{\parallel}=1$; (b) $\Delta/J_{\parallel}=2$.}
\label{fig:MagFullHeis}
\end{figure*}

It can be concluded that the intermediate magnetization plateaus at zero and one-half of the saturation magnetization for $\Delta/J_{\parallel}=1$ (see Figs. \ref{fig:Mag_dmrgLancz}(a)-(d)), and at zero, one-quarter and one-half of the saturation magnetization for $\Delta/J_{\parallel}=2$ (see Figs. \ref{fig:Mag_dmrgLancz}(e)-(h)) represent the actual magnetization plateaus. To complete our discussion, we study the effect of four-spin Ising coupling on the magnetization process of the spin-1/2 Heisenberg model on the double sawtooth ladder. 3D zero-temperature plot of the magnetization of the spin-1/2 Heisenberg double sawtooth ladder is presented in Fig. \ref{fig:MagFullHeis}as a function of the magnetic field $B/J_{\parallel}$ and the cyclic four-spin Ising interaction $K/J_{\parallel}$.

Figure \ref{fig:MagFullHeis}(a) displays the magnetization normalized with respect to its saturation value as a function of the magnetic field and four-spin Ising interaction for fixed values of the interaction parameters $\eta/J_{\parallel}=J/J_{\parallel}=1$ and $\Delta/J_{\parallel}=1$. It is worthwhile to recall that the magnetization curve displays for $K/J_{\parallel}=0$ two actual intermediate magnetization plateaus at zero and one-half of the saturation magnetization (based on our DMRG results shown in Fig. \ref{fig:Mag_dmrgLancz}). With increase of the four-spin coupling $K/J_{\parallel}$, one observes substantial changes of the magnetization curve with a gradual shrinking of zero and one-half magnetization plateaus, whereas the intermediate one-quarter plateau contrarily becomes broader. It is quite plausible to conjecture that there are accordingly strong indications for appearance of the one-quarter and three-quarter magnetization plateaus for nonzero values of the Ising four-spin interaction $K/J_{\parallel}$. In opposite to this, the intermediate one-half magnetization plateau may terminate for nonzero $K/J_{\parallel}$ at a special quantum critical point.
 If we turn back to Fig. \ref{fig:GSPD_BK}(b), we find out that similar scenario happens also for the one-half magnetization plateau of the spin-1/2 Ising-Heisenberg double sawtooth ladder corresponding to the ground state $\vert\text{nunu}\rangle$, which gradually shrinks upon increasing of the four-spin coupling $K/J_{\parallel}$ and ultimately terminates at a critical point with the coordinates  $[K/J_{\parallel}, B/J_{\parallel}]\approx [0.8, 3.4142]$.

Even more striking features can be figured out in the magnetization curves plotted in Fig. \ref{fig:MagFullHeis}(b) for $\Delta/J_{\parallel}=2$. Based on the DMRG results shown in Fig. \ref{fig:Mag_dmrgLancz} it is obvious that the magnetization curve of the spin-1/2 Heisenberg double sawtooth ladder exhibits for $\Delta/J_{\parallel}=2$ and $K/J_{\parallel}=0$ three intermediate plateaus at zero, one-quarter and one-half of the saturation magnetization together with three gapless spin-liquid regions separating them. The full ED calculations unveil that the intermediate one-half magnetization plateau becomes narrower upon increasing of the four-spin Ising interaction $K/J_{\parallel}$.
By inspecting Fig. \ref{fig:GSPD_BK}(b), one can realize that in the ground-state phase diagram of the spin-1/2 Ising-Heisenberg double sawtooth ladder the area of the ground state $\vert\text{nunu}\rangle$($\vert\text{usus}\rangle$) corresponds to the one-half magnetization plateau and it also gradually decreases upon increasing of the four-spin coupling $K/J_{\parallel}$ until it disappears at a quadruple point $[K_\text{Q}/J_{\parallel}, B_\text{Q}/J_{\parallel}]\approx [0.2929, 4.4142]$. On the other hand, the parameter region corresponding to the other ground state $\vert\text{nuuu}\rangle$ related to the three-quarter magnetization plateau increases upon increasing of the four-spin interaction term $K/J_{\parallel}$.

\section{Conclusions}\label{conclusion}
In this work, we have exactly solved the spin-1/2 Ising-Heisenberg double sawtooth ladder supplemented with the four-spin Ising interaction by taking advantage of the classical transfer-matrix technique. The ground-state phase diagram, the magnetocaloric properties and magnetic Gr{\" u}neisen parameter of this model were rigorously examined. We found a peculiar quadruple point in the ground-state phase diagram at which four different ground states coexist together, whereby this quadruple point cannot be observed for the pure Ising double sawtooth ladder. Besides, the enhanced MCE has been detected nearby the triple and quadruple points. The exact results for the isentropes of the spin-1/2 Ising-Heisenberg double sawtooth ladder evidenced a fast cooling during the adiabatic demagnetization process when the magnetic field is tuned to a close vicinity of the triple and quadruple points. The coordinates of the quadruple point depends on a relative ratio between the isotropic exchange interaction between the Heisenberg dimers and the nearest-neighbor Ising interaction along the legs. For a particular value of the four-spin Ising interaction that drives the investigated spin-1/2 Ising-Heisenberg model to a quadruple point one even detects  the enhanced MCE at relatively high temperatures.

By employing the numerical ED and DMRG methods we also investigated the magnetization process of the spin-1/2 Heisenberg double sawtooth ladder with and without the four-spin Ising coupling. A proper finite-size analysis allowed us to discern the true intermediate magnetization plateaus from the gapless quantum spin-liquid regions.
The ED data imply that the intermediate one-half magnetization plateau gradually shrinks upon increasing of the four-spin Ising coupling  and it seems to vanish at a special quantum critical point.
A gradual suppression of the gapped one-half plateau phase due to the rising four-spin Ising interaction, which possibly closes an energy gap, may be an indication of the Kosterlitz–Thouless quantum critical point. The precise nature of the quantum critical point can be an interesting subject of future studies.

\section*{Acknowledgments}
H. Arian Zad acknowledge the financial support of the National Scholarship Programme of the Slovak Republic (NŠP).
V. Ohanyan acknowledges partial financial support form ANSEF (Grants No. PS-condmatth-2462 and PS-condmatrth-2884), J. Stre\v{c}ka and A. Zoshki acknowledge the financial support by the grant of the Slovak Research and Development Agency provided under the contract No. APVV-20-0150 and by the grant of The Ministry of Education, Science, Research, and Sport of the Slovak Republic provided under the contract No. VEGA 1/0105/20.


\bibliography{}

\end{document}